\documentclass[onecolumn,secnumarabic,amssymb, nobibnotes, aps,notitlepage, pr,superscriptaddress,10pt]{revtex4-2}

\usepackage{amsmath,amssymb}
\usepackage{stackengine,graphicx}

\graphicspath{{images/}} 
\usepackage[dvipsnames]{xcolor}
\usepackage{bbm}
\usepackage{enumitem}
\usepackage{colortbl}
\usepackage{lipsum}

\usepackage{tabu}
\usepackage{makecell}

\usepackage{xurl}

\PassOptionsToPackage{hyphens}{url}\usepackage{hyperref}

\usepackage{subcaption}

\usepackage{soul}

\usepackage{collectbox}
\usepackage{bbm}

\UseRawInputEncoding
\usepackage[utf8]{inputenc}

\usepackage{hyperref}
\hypersetup{
    colorlinks=true,
    linkcolor=blue,
    filecolor=magenta,      
    urlcolor=cyan,
    citecolor=black
}

\def\l{\left}
\def\r{\right}

\newcommand{\f}{\frac}

\usepackage{array}
\newcolumntype{L}[1]{>{\raggedright\let\newline\\\arraybackslash\hspace{0pt}}m{#1}}
\newcolumntype{C}[1]{>{\centering\let\newline\\\arraybackslash\hspace{0pt}}m{#1}}
\newcolumntype{R}[1]{>{\raggedleft\let\newline\\\arraybackslash\hspace{0pt}}m{#1}}

\usepackage{accents}

\newcommand{\lu}{College of Health, Lehigh University, Bethlehem, Pennsylvania, United States of America}

\newcommand{\metaculus}{Metaculus, Santa Cruz, California, United States of America}

\newcommand{\gjo}{Good Judgement Inc., New York, New York, United States of America}

\date{\today} 

\usepackage[paperwidth=8.5in,paperheight=11.0in,
  left=1.in,right=1.in,top=1.5in,bottom=1.5in,
  includefoot,heightrounded]{geometry}

\begin{document}

\title{Chimeric forecasting: combining probabilistic predictions from computational models and human judgment}

\author{Thomas~McAndrew}
\email{mcandrew@lehigh.edu}
\affiliation{\lu}

\author{Allison~Codi}
\affiliation{\lu}

\author{Juan~Cambeiro}
\affiliation{\metaculus}

\author{Tamay~Besiroglu}
\affiliation{\metaculus}

\author{David~Braun}
\affiliation{\lu}

\author{Eva Chen}
\affiliation{\gjo}

\author{Luis Enrique Urtubey de C\`{e}saris}
\affiliation{\gjo}

\author{Damon~Luk}
\affiliation{\lu}

\setlength{\parindent}{0em}
\setlength{\parskip}{0.5em}

\begin{abstract}
Forecasts of the trajectory of an infectious agent can help guide public health decision making.
A traditional approach to forecasting fits a computational model to structured data and generates a predictive distribution.
However, human judgment has access to the same data as computational models plus experience, intuition, and subjective data. 
We propose a chimeric ensemble---a combination of computational and human judgment forecasts---as a novel approach to predicting the trajectory of an infectious agent.
Each month from January, 2021 to June, 2021 we asked two generalist crowds, using the same criteria as the COVID-19 Forecast Hub, to submit a predictive distribution over incident cases and deaths at the US national level either two or three weeks into the future and combined these human judgment forecasts with forecasts from computational models submitted to the COVID-19 Forecasthub into a chimeric ensemble.
We find a chimeric ensemble compared to an ensemble including only computational models improves predictions of incident cases and shows similar performance for predictions of incident deaths.
A chimeric ensemble is a flexible, supportive public health tool and shows promising results for predictions of the spread of an infectious agent.
\end{abstract}

\maketitle

\clearpage

\section{Introduction}

Forecasts of the transmission and burden of COVID-19 provide public health officials advance warning that allows them to make informed decisions about how to modify their response to the pandemic~\cite{lutz2019applying,biggerstaff2021improving,biggerstaff2020early,hufnagel2004forecast}.
The COVID-19 pandemic has caused economic burdens to the US, overwhelmed hospitals with ill patients, and further highlighted social inequity and inequalities in access to healthcare~\cite{di2021health,chen2021economic,khullar2020covid,kaufman2020half,mehrotra2020personal,wu2021hospital}.

In response, several organized modeling efforts were started to give public health officials as up to date information as possible about the trajectory of COVID-19 in the US and in Europe~\cite{ray2020ensemble,borchering2021modeling,bracher2021pre,europeancovid}.

The US COVID-19 Forecast Hub is a unified effort to house probabilistic forecasts of incident cases, deaths, and hospitalizations due to COVID-19 in a single, centralized repository~\cite{ray2020ensemble,cramer2021united}.
The goal of this repository is to collect, combine, and evaluate forecasts of the trajectory of COVID-19 and communicate these forecasts to the public and to public health officials at the state and federal level~\cite{cramer2021evaluation}.
This repository is not meant to include all possible forecasting targets related to COVID-19, and models not included in the COVID-19 Forecast Hub have forecasted vaccine safety, efficacy, and timing, conditional trajectories of COVID-19 given public health action, time-varying $R_{0}$ values, hospital bed requirements, among others~\cite{mcandrew2021aggregating,abbott2020estimating,goic2021covid,ferstad2020model,papastefanopoulos2020covid,maleki2020time,ingle2021projecting}. 
The strength of the COVID-19 Forecast Hub is it's ability to store, evaluate and communicate forecasting efforts systematically and focus modeling efforts that process objective, reportable data.

In addition to the US COVID-19 forecast hub, there are COVID-19 hubs that collect computational forecasts for Europe and specifically for Germany and Poland~\cite{ray2020ensemble,bracher2021pre,europeancovid}. 
The majority of models submitted to these hubs are computational: statistical or dynamical models trained on structured data.

Statistical models build a forecast by leveraging correlations between the current trajectory of COVID-19 and a set of covariates~\cite{gnanvi2021reliability,jewell2020predictive,meehan2020modelling,latif2020leveraging,guan2020modeling,arik2020interpretable,shinde2020forecasting,mac2021modeling,dimitrov2010mathematical,grassly2008mathematical}. 
Traditional data sources that were used to train models include historical counts of incident cases, deaths, and hospital admissions. 
A subset of models also train on novel sources of data such as self-reported COVID symptom rates and the rate of visits to a doctor, data related to mobility or contact among individuals, and social media data~\cite{wilson2021weather,reinhart2021open,lee2020human,shen2020using}.

Dynamical models first pose a deterministic relationship for how an outbreak is expected to evolve and then typically add a random variable to account for uncertainty between the (conjectured to be true) deterministic process and what is reported~\cite{tolles2020modeling,weiss2013sir,kermack1927contribution}.
The most common dynamical models of the trajectory of COVID-19 extend compartmental models, models that assume individuals are in one of a finite set of states through the pandemic, to incorporate time varying reproduction numbers, multiple different data sources, and more complicated spatial structure ~\cite{yang2021time,chen2020time,aleta2020modelling,gibson2020real}. 
Dynamical models often excel at long term forecasts and generating a conditional probability over an epidemiological variable in response to public health action~\cite{liu2020forecasting,fowler2020effect,pei2020differential,kudryashov2021analytical,chinazzi2020effect,aleta2020modelling}.

Human judgment forecasting relies on the beliefs and activities of a crowd to generate (point or probabilistic) predictions over the possibilities of some future event.

Prediction markets have been developed to predict infectious diseases such as the 2009 swine flu, seasonal influenza, enterovirus, and dengue fever~\cite{ritterman2009using,li2016wisdom,tung2015using}.
A prediction market provides participants an initial amount of "money" to spend on future events and allows participants to place higher bids on events they think are more likely to occur. 
After bidding is complete, a model maps the "market price" for each event to a probability which is interpreted as the crowd's belief that event will occur~\cite{wolfers2004prediction}.
Prediction markets rely on a large and diverse participant pool and the model that connects market price to predictive probability to make accurate predictions~\cite{kambil2002making,mchugh2012prediction}.

Passive human activity and behavior from social media 
outlets like Twitter and Facebook, and internet search history have been used as inputs to a model and have shown improved accuracy compared to a model that uses only epidemiological data for infectious agents like influenza, dengue fever, ZIKA, and COVID-19~\cite{samaras2020comparing,al2016using,alessa2018review,masri2019use,marques2017dengue,ning2019accurate}.
Most models (i) extract features from these social media outlets, (ii) transform the extracted social media data and include objective epidemiological data, and (iii) train a predictive model on this combination of objective, subjective data.
Models using social media data are usually statistical or machine learning models, exploiting correlations between these data sources and the target of interest.

Direct predictions---either point predictions or probability densities---of the trajectory of an infectious agent have been elicited from individuals and aggregated for diseases such as influenza and COVID-19~\cite{mcandrew2020expert,mcandrew2021aggregating,farrow2017human,bosse2021comparing}.
Point forecasts have been elicited from experts from platforms like Epicast~\cite{farrow2017human}.
Epicast asks participants to predict the entire trajectory of influenza-like illness~(ILI), a marker for the severity of seasonal influenza, by viewing the current ILI time series and then drawing a proposed trajectory from the present week to the end of the influenza season.
The aggregate model assigns a probability to an ILI value belonging in the bounded interval $[x,x+\delta]$ as the proportion of individual trajectories that fall within those bounds.
The Epicast model was routinely one of the top performing models among several computational models submitted to the CDC sponsored FluSight challenge~\cite{farrow2017human}.

Three projects to date have collected direct, probabilistic predictions from humans about the transmission and burden of the COVID-19 pandemic~\cite{mcandrew2020expert,recchia2021well,bosse2021comparing}. 
As early as February 2020, human judgment platforms have made predictions of the trajectory of COVID-19 by enrolling experts in the modeling of infectious disease and asking them questions related to reported and true transmission, hospitalizations, and deaths due to SARS-CoV-2~\cite{mcandrew2020expert}.
Experts were also asked to make predictions of transmission conditional on future public health actions.
An equally weighted average of expert predictions was used to combine individual predictions into consensus predictions and reports from this work were generated from February 2020 to May 2020. 
This work found that, although there was considerable uncertainty assigned to confirmed cases and deaths, a consensus of expert predictions was robust to poor individual predictions, able to make accurate predictions of confirmed cases one week into the future, and gave an early warning signal of the severity of SARS-CoV-2.
The second project compared predictions of rates of infection and number of deaths between those who were considered experts and laypeople in the United Kingdom~\cite{recchia2021well}.
Participants were asked to assign a 12.5th and 87.5th percentile to four questions related to COVID-19---one question with ground truth and three with estimated values for the truth. 
Expert predictions were more accurate and calibrated than non-expert predictions, however expert predictions still underestimated the impact of COVID-19.
A third project solicited from experts in statistics, forecasting, and epidemiology direct predictions of one through four week ahead incident and cumulative cases and deaths for Germany and Poland (at the national level) and aggregated these predictions into a "crowd forecast"~\cite{bosse2021comparing}.
The crowd was able to contribute to a more predictive forecast of cases in both countries, however predictions of deaths may have not added much predictive accuracy. 

Human judgment predictions have been applied to a numerous number of fields beyond infectious disease and interested readers can find comprehensive reviews on the status and applications of human judgement forecasting~\cite{mcandrew2021aggregating,hanea2021uncertainty,clemen1989combining}. 
Select foundational works on aggregating human judgment may be found in the following citations~\cite{bates1969combination,clemen1989combining,clemen1999combining,winkler1989combining,genest1990allocating}.  

To the best of our knowledge, we propose the first ensemble algorithm designed to generate forecasts of the trajectory of an infectious agent by combining direct, probabilistic predictions from computational models and human judgement models.
We call this type of ensemble a \textit{chimeric ensemble}.
In this first hypothesis-generating work we: (i) explore the advantages and challenges when combining computational and human judgment models, (ii) compare the performance of a chimeric ensemble to a computational model only ensemble on six forecasts of incident cases and six forecasts of incident deaths due to COVID-19 at the US national level between January 2021 and June 2021, and (iii) compare and contrast an algorithm that differential weights models based on past performance to an equally weighted combination of models.

\section{Methods}

\subsection{Forecasting logistics}

\subsubsection{Survey timeline}

Six monthly surveys were sent to experts and trained forecasters from January to June 2021 on the Metaculus forecasting platform~\url{https://www.metaculus.com/} and five monthly surveys from February to June 2021 were sent to the Good Judgment Open (GJO) platform~\url{https://www.gjopen.com/}. 
Participants had approximately ten days to add probabilistic predictions, and were encouraged to include a rationale alongside their quantitative forecasts to provide insight into how they made their predictions.
Participants on both platforms were allowed to revise their original predictions as many times as they wished between when the survey was open and when it closed (often ten days later).
During the course of all six surveys, participants could revisit their past predictions but could no longer revise predictions for those surveys that were closed.
A list of survey open and close times, questions that were asked, and how the truth was determined for each question can be found in supplement~\ref{tab.quesres}. 

The Lehigh University Internal Review Board determined that this work does not meet the definition of human subjects research.

\subsubsection{Forecaster elicitation}

All subscribers to the Metaculus platform and to the GJO platform were invited to make anonymous predictions of epidemiological targets related to COVID-19.
Subscribers to Metaculus were sent email invitations and all questions related to this project were grouped together and posted on the Metaculus website as a tournament titled \textit{Consensus Forecasting to Improve Public Health: Mapping the Evolution of COVID-19 in the U.S.} which can be found at \url{https://pandemic.metaculus.com/questions/?search=contest:consensus--forecasting}. 
Subscribers to GJO were invited to participate via email and questions for this project were posted on the GJO website as "Featured Questions".
A convenience sample of $16$ experts were invited to participate on the Metaculus platform.
We defined an expert as one who has several years of experience in the study or modeling of infectious disease and have kept up to date on scientific literature, and public health efforts related to COVID-19. 

Both the Metaculus and GJO platforms offer training and prediction resources on their websites~(\url{https://www.metaculus.com/help/prediction-resources/} and \url{https://goodjudgment.com/services/online-training/}) that allows a subscriber to familiarize themselves (i) with how to make calibrated and accurate predictions and (ii) how to use the tools and features of the platform. 

Forecasters on Metaculus and Good Judgment receive, for each question they answer on the website, immediate feedback from a visualization of the present consensus forecast and longer term feedback by receiving an email when the ground truth for a question resolves and a score that determines the accuracy of their prediction for a specific question. 

\subsubsection{How predictions were collected from humans}

Forecasters submitted monthly predictions in a format that depended on if they used the Metaculus platform or the Good Judgment open platform.

Participants on Metaculus generate predictions over a continuous bounded interval as a combination of up to five logistic distributions~(Supplemental~Fig.~\ref{fig.metac}).
When a participant decides to form a prediction they are presented with a single logistic distribution and a slider bar underneath this distribution.
The slider contains a square indicating the distribution median and two circles to the left and right of the square that help identify the distribution's 25th and 75th quantiles. 
Participants can shift this distribution left, over smaller values, or right, over larger values, by moving the square and they can scale this distribution by expanding or contracting the circles to the left and right of the square.
If a participant decides to include a second (third, fourth, and fifth) logistic distribution they can select "add component".
A second predictive density is overlaid over the first and the participant can control that second density by using a second slider that appears below the first.
In addition to the two sliders, an additional two slider bars appear that allow the participant to assign weights to the first and second (third, fourth, fifth) predictive densities. 

Participants on GJO assign probabilities to a set of intervals $I_{1},I_{2},\cdots,I_{n}$ that partition an open interval~(Supplemental~Fig.~\ref{fig.gjo}).
For each interval $I_{i}$, participants are presented a slider bar controlling the probability assigned to $I_{i}$ and that can be at minimum zero and maximum one. 
To the right of each slider bar is a text box that contains the current probability the participant has assigned to $I_{i}$. 
The probabilities assigned to all intervals must sum to one, and as a participant selects probabilities to assign to each interval the  total probability is computed and displayed.
A participant can only submit a probability distribution when the total probability equals one.

\subsection{COVID-19 Forecast Hub}

The COVID-19 Forecast Hub collects prospective forecasts of the trajectory of COVID-19 in the United States from more than 80 computational models~\cite{Cramer2021-hub-dataset,cramer2021evaluation,ray2020ensemble}.
Forecasts of weekly incident cases are produced at the national, state, and county level, and forecasts of weekly incident and cumulative deaths and daily hospitalizations are produced at the national and state levels. 
Forecasts of cases are submitted to the COVID-19 Forecast Hub as a set of 7 quantiles and forecasts of deaths are submitted as a set of 23 quantiles.
Models produce predictions of weekly cases and deaths one, two, three, and four weeks ahead.
A GitHub repository~(\url{https://github.com/reichlab/covid19-forecast-hub})~is used to keep track of individual submissions and an ensemble model.

\subsection{Human judgement forecasting targets}

Members of the Metaculus and GJO crowd were asked to predict the number of incident cases and incident deaths due to COVID-19 that would be observed at the US national level over the course of one epidemic week. 
These "core" questions were asked for all six surveys, were presented to humans in the same format for all six surveys, and were meant to match, as much as possible, to the corresponding forecast targets used by the COVID-19 Forecast Hub.

In addition to these core questions, we asked the Metaculus crowd only extra questions of public health relevance.
Example questions include the cumulative number of first and full dose vaccinations by a given date, cumulative deaths by year end, the 7-day moving average of the percent of B.1.1.7 in the US, and the incident number of weekly hospitalizations. 
A list of all questions asked throughout the six surveys can be found in the supplement~(Supplemental~\ref{tab.quesres}).

\subsection{Matching COVID-19 Forecast Hub and human judgement forecasting targets}

How questions were posed to human judgement crowds and how the truth was determined for questions related to incident cases and incident deaths at the US national level matched how the ground truth was determined by the COVID-19 Forecast Hub. 
When we described the resolution criteria for forecasts of incident cases and deaths, we matched, as close as possible, the ground truth document sent to modeling teams who submit computational forecasts to the COVID-19 Forecast Hub~(technical readme for COVID-19 Forecast Hub: \url{https://github.com/reichlab/covid19-forecast-hub/blob/master/data-processed/README.md}).

The COVID-19 Forecast Hub allows computational forecasts to be submitted at any time, but only computational forecasts that are submitted on Mondays of each week are included in the weekly COVID-19 forecast hub ensemble. 
Each survey sent to Metaculus and GJO crowds was open for submission before a COVID-19 Forecast Hub due date. 
In January surveys closed six days after the Monday due date, in February and March surveys closed on a Monday deadline, in April and May surveys closed one day after a COVID-19 Forecast Hub due date, and in June two days after a due date.
Individual predictions submitted to Metaculus and to GJO were cut at the same due date as the one asked of computational models submitted to the COVID-19 Forecast Hub
~Fig.~\ref{fig.experimentAndDataCollection}~(A.).
Counts of the number of computational and human judgement models can be found in supplemental~\ref{tab.counts}.
The goal with cutting individual predictions at the same time as computational model was for a fair comparison, and a fair combination of computational and human judgement forecasts.

\subsection{Data availability and real-time summary reporting}
Human judgement consensus predictions and chimeric predictions of incident cases and incident death using an equally weighted ensemble approach are available for all surveys at the Zoltar Forecast Archive: \url{https://zoltardata.com/model/511}. 
Anonymized data on individual predictions is available upon request.

Summary reports that were generated in real-time from January 2021 to June 2021 on all targets (not just cases and deaths) are available at \url{https://github.com/computationalUncertaintyLab/aggStatModelsAndHumanJudgment_PUBL}.

\subsection{Forecast scoring}
\label{scoring}

\def\IS{\text{IS}}
\def\WIS{\text{WIS}}
\def\CRPS{\text{CRPS}}

Individual, consensus, ensemble, and chimeric forecasts were scored using the weighted interval score (WIS) over $K$ central quantiles~\cite{bracher2021evaluating}.
\begin{equation*}
\WIS_{\alpha_{\{0:K\}}}(F,y) = \frac{1}{K + 1/2} \l( w_0 \times |y-m| + \sum_{k=1}^{K} \{w_k \times \IS_{\alpha_k}(F,y)\} \r) 
\end{equation*}
where the interval score $(\IS_{\alpha_{k}})$ is 
\begin{equation*}
    \IS(\alpha)(F,y) = (u-l) + \f{2}{\alpha} (l-y) \mathbbm{1}(y<l) + \f{2}{\alpha} (y-u) \mathbbm{1}(y>u)
\end{equation*}
and where $F$ is a predictive cumulative distribution function, $\mathbbm{1}(x)$ is an indicator function, the value $u$ represents the $(1-\alpha/2)$ quantile of $F$, $l$ represents the $\alpha/2$ quantile of $F$, and $m$ represents the median or 0.50 quantile, and $y$ is eventually reported truth~\cite{gneiting2007strictly}.
Weight $w_{0}$ equals $\f{1}{2}$ and $w_{k} = \f{\alpha_{k}}{2}$.

The weighted interval score (and interval score) are negatively sensed---larger values indicate worse predictive performance compared to smaller values. 
The best possible weighted interval score is zero and the worst possible weighted interval score is positive infinity.

WIS is a discrete approximation of the continuous rank probability score
\begin{equation*}
    \CRPS(F,y)= \int_{-\infty}^{\infty} \{F(x) - 1(x \geq y)\}^2 \,dx 
\end{equation*}
where the WIS score converges to the same value as the $\CRPS$ as the number of equally spaced intervals $(K)$ increases given a fixed cumulative density $F$ and true value $y$ ~\cite{bracher2021evaluating}.

The WIS is the score adopted by the Centers for Disease Control and Prevention~(CDC) to evaluate forecasts of incident cases, deaths, and hospitalizations submitted as a set of set of central quantiles.

The WIS and CRPS are examples of negatively sensed proper scoring rules~\cite{gneiting2007strictly,gneiting2007probabilistic}.
A negatively sensed proper scoring rule is a function $S$ that takes as input a density $F$ and true value $y$ and returns a non-negative real number that is minimized when the input density $F$ is distributed the same as the true data generating process $Y$ that produced the true, realized value $y$~\cite{gneiting2007strictly,gneiting2007probabilistic}. 

\subsection{Consensus algorithm strategies}

\subsubsection{Data setup}

Ensemble forecasting of infectious targets involves three related data sets: (i) data collected about epidemiological quantities of interest, $\mathcal{D}$, (ii) predictive densities over these targets submitted by individual models (either computational or human), $F$, and (iii) a score given to each model forecast about a collected data point, $\mathcal{S}$. 

We suppose an epidemiological target, or quantity of interest~(incident cases, deaths, etc.) at time $t$ can be represented by a random variable $T_{t}$, and further assume true values $\mathcal{D} = [t_{1},t_{2},\cdots,t_{N}]$ were generated by random variables $T_{1}, T_{2}, \cdots, T_{N}$ where $T_{t}$ is specific to a single target, point in time, and location. 
We make no additional assumptions about whether targets are dependent or independent and do not assume a specific distribution over potential target values. 

A model produces a forecast for a target $T_{t}$ in the form of a set of $K$ centralized quantiles.
We can organize forecasts $F$ over all targets from $M$ models that submitted $K$ quantiles into a matrix where a forecast from a single model corresponds to one row and one column corresponds to a quantile about one target.
For example, a forecast matrix with $3$ models, $K$ quantiles, and $T$ targets can be formed as follows
\begin{equation*}
    F = \begin{bmatrix}
            \text{\underline{Model}} & \multicolumn{4}{c}{\underline{\text{Target 1}}} & \multicolumn{4}{c}{\text{\underline{Target 2}}} & \hspace{3mm} \cdots & \multicolumn{4}{c}{\text{\underline{Target T}}} \vspace{0.75mm} \\
            M_{1} |  & q^{1}_{1,1} & q^{1}_{1,2} &  \cdots & q^{1}_{1,K} & \hspace{3mm} q^{1}_{2,1} & q^{1}_{2,2} & \cdots & q^{1}_{2,K} & \hspace{3mm} \cdots  & \hspace{3mm} q^{1}_{T,1} & q^{1}_{T,2} & \cdots & q^{1}_{T,K}    \vspace{0.75mm}\\ 
            M_{2} |  & q^{2}_{1,1} & q^{2}_{1,2} & \cdots & q^{2}_{1,K} & \hspace{3mm} q^{2}_{2,1} & q^{2}_{2,2} & \cdots & q^{2}_{2,K} & 
            \hspace{3mm} \cdots & \hspace{3mm} q^{2}_{T,1} & q^{2}_{T,2} & \cdots & q^{2}_{T,K} \vspace{0.75mm}\\  
             M_{3} |  & q^{3}_{1,1} & q^{3}_{1,2} & \cdots & q^{3}_{1,K} & \hspace{3mm} q^{3}_{2,1} & q^{3}_{2,2} & \cdots & q^{3}_{2,K}  &
             \hspace{3mm} \cdots & \hspace{3mm} q^{3}_{T,1} & q^{3}_{T,2} & \cdots & q^{3}_{T,K} \\ 
        \end{bmatrix}
\end{equation*}
No assumptions about a predictive density are placed on models beyond requiring a list of $K$ quantile values.

A matrix $\mathcal{S}$ can also be built
\begin{equation*}
    \mathcal{S} = 
        \begin{bmatrix}
            s_{1,1} & s_{1,2} & \cdots & s_{1,N}\\ 
            s_{2,1} & s_{2,2} & \cdots & s_{2,N} \\
            \vdots  &         & \ddots & \vdots \\
            s_{M,1} & s_{M,2} & \cdots & s_{M,N} \\
         \end{bmatrix}
\end{equation*}
where the $\mathcal{S}_{ij}$ entry of this matrix, $s_{ij}$, corresponds to the score for model $i$ about target $j$

\subsubsection{Model combination and optimization}

We chose to combine individual forecasts for our consensus and chimeric ensembles using a quantile average.
We define a quantile average as a convex combination of individual forecast quantiles
\begin{equation*}
    f = F' \pi
\end{equation*}
where $f$ is a row vector of length $KN$ and $\pi = [\pi_{1},\pi_{2},\cdots,\pi_{M}]$ is a vector of length $M$.
The weight vector $\pi$ is further constrained to have non-negative entries and to sum to one.

We will estimate weights for each model by finding a vector $\pi$ such that the ensemble forecast $f$ minimizes in-sample mean WIS scores $(W)$ over all targets with ground truth available.
Given a sample of $T$ realized true values $\mathcal{D} = [t_{1},t_{2},\cdots,t_{T}]$ 
\begin{equation}
\begin{aligned}
    &\min_{f} \overline{ W(f)} \; \; \text{s.t.}\\
    &\pi' \mathbbm{1} = 1 \\
    & 0 \le \pi_{m} \le 1 \label{optim}
\end{aligned}
\end{equation}
where $\mathbbm{1}$ is a vector of ones, $W(f)$ is a vector of WIS scores for $f$, and $\overline{W(f)}$ is the average WIS score for an ensemble density $f$ over all targets.
Because we choose weights $\pi$ to assign to out of sample probabilistic predictions which minimize an objective function, this process can be framed as a specific case of stacked generalization~\cite{wolpert1992stacked}.

The algorithm we chose to optimize the weights assigned to computational and human judgment models is a population based optimization strategy called differential evolution.
Differential evolution~(DE) is a stochastic direct search method that is often robust to high dimensional parameter spaces and multi-modal objectives~\cite{storn1997differential}. 

Given a set of $M$ computational and human judgment forecasts at survey time $T$, the goal of this DE algorithm is to find a $M \times 1$ vector used to weight individual models that minimizes the mean WIS over all past survey time points for which we have the truth.
To begin, DE chooses at random $4$ $M \times 1$ vectors and evaluates the  mean WIS score for each of the four weight vectors.
At the next iteration each of the potential vector solutions, in turn, is compared to a new candidate vector solution.
The candidate vector solution to be compared is generated by "mutation" and "crossover"~(details can be found in ~\cite{storn1997differential}).
Mutation and cross over have associated parameter values, and we chose a value of $0.8$ for mutation and $0.9$ for cross over.
If the candidate solution reports a smaller mean WIS score than the original vector, the original vector is replaced with this new solution. 
This iteration is complete after all original solutions have been compared to new candidate solutions.
Then the next iteration starts.
All solutions were normalized by dividing the $M \times 1$ potential vector solution by the sum of all the entries to guarantee the final, minimal solution assigned weights that sum to one. 
Differential evolution was implemented by using the python package \textbf{mystic}~\cite{mckerns2012building,mckernsmystic}.

\subsubsection{Methods to account for missing forecasts}

We took three approaches to impute missing forecasts: (i) a complete case approach, (ii) an available forecast approach we call "spotty memory", and (iii) an approach we call "defer to the crowd".

The complete case approach combines models that have made forecasts for all targets asked for the present survey and all past surveys.
If a model missed a forecast, past or present, they are removed from the ensemble.
The "spotty memory" approach combines models if they have made forecasts for all targets in the present survey.
If a model missed a forecast in the past they are still included.
If a model missed a forecast for the present survey for either cases or deaths than they are removed from the ensemble.
The "defer to the crowd" approach combines models that have made at least one forecast for any past or present survey. 
A model without a forecast for the present survey is included.

The complete case approach will have no missing forecasts, however we must impute missing forecasts for both the "spotty memory" and "defer to the crowd" approach.
To impute a missing forecast, we considered each quantile a function of $K$ quantiles submitted by $M$ models about a single target. 
We only allow predictions of the same target to inform missing forecasts. 

Define a matrix $Q$  by selecting only those quantiles from $F$ that correspond to a single target.
The rows of $Q$ correspond to models and the columns correspond to $K$ quantiles where the smallest quantile is the first column, the second smallest quantile is the second column, up until quantile $K$.
We denote $Q_{-k}$ as the matrix $Q$ with column $k$ removed and $Q_{k}$ as the $k^{\text{th}}$ column vector of $Q$.

Then we can impute $Q_{k}$ as a function $g$ which takes as input $Q_{-k}$ and potentially some parameter set $\theta$
\begin{align*}
    Q_{k} = g(Q_{-k},\theta)
\end{align*}

We chose to test the following 5 approaches to impute missing forecasts: mean imputation, median imputation, bayesian ridge regression, decision tree regression and extremely randomized trees~(see table~\ref{tab.impute} for a summary of these methods). 

For the last three regression approaches, missing quantiles were imputed using a chained equation process. The chained equation process imputed missing values in four steps.
Step one, replace missing quantiles in $Q_{k}$ with the mean over all present quantiles in column $k$.
Step two, choose the column with the fewest missing values, set the values imputed with the mean back to missing.
Step 3, impute missing values for column $k$ using $ g(Q_{-k},\theta)$.
Step 4, repeat the above process on the quantile with the second fewest number of missing values.
The above steps are iterated until convergence.
We used the "IterativeImputer" function from scikit-learn to perform this chained equation imputation~\cite{pedregosa2011scikit}.

\section{Results}

\subsection{Survey logistics and participation}

A total of six surveys were run from January 2021 to June 2021. 
Each survey asked on average $7.5$ questions related to national level incident cases, incident deaths, incident hospitalizations, the cumulative number of first dose and fully vaccinated individuals, and additional questions of immediate public health importance such as the proportion of sequences classified as B.1.17 among all sequenced viruses. 
A list of all questions asked for each survey can be found in~supplemental section~\ref{tab.quesres}. 
At the end of each month a summary report was generated and posted online~(summary reports can be found at the following link=\url{https://github.com/computationalUncertaintyLab/aggStatModelsAndHumanJudgment_PUBL}). 

We collected from the Metaculus platform predictions from 68 unique members who made a total of 1062 original and revised predictions across all twelve questions related to cases and deaths.
From GJO we collected predictions from 323 unique members who made 3319 original and revised predictions. 

From the COVID-19 Forecast Hub we collected a total of 364 predictions of incident cases and incident deaths at the national level generated by 46 computational models between January and June of 2021.
Computational models used a variety of techniques to build predictions of incident cases and deaths such as traditional statistical time series models like ARIMA and state space models, machine learning techniques such as deep artificial neural networks, and compartmental models.
A list of the computational models included in this analysis can be found in supplement~\ref{tab.listOfCompModels}.

\begin{figure}[ht!]
    \centering
    \includegraphics{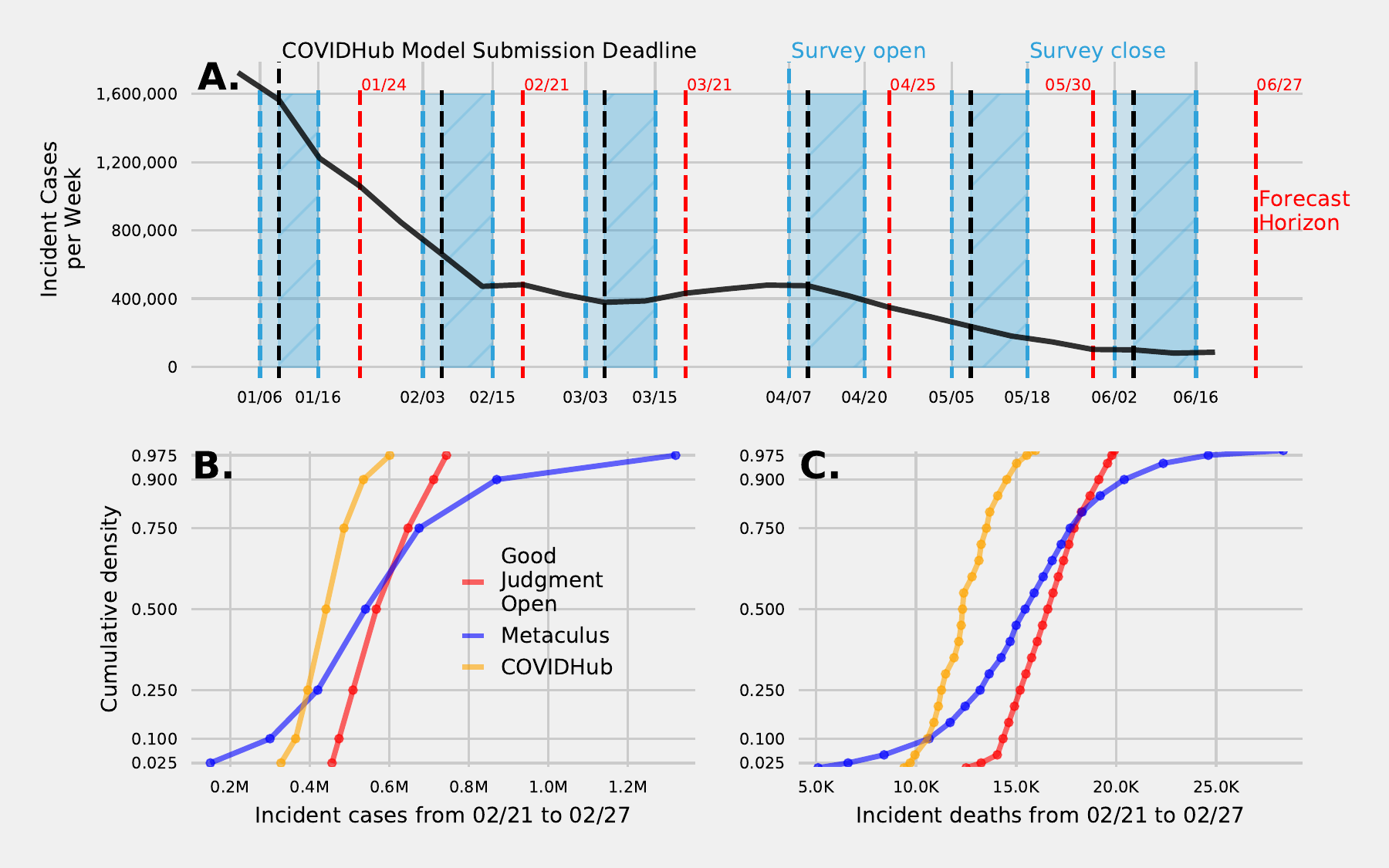}
    \caption{
    (A.)~A timeline of the six surveys that collected human judgment predictions from January to June of 2021, showing when surveys were open and closed (blue dashed lines), when computational predictions submitted to the COVID-19 Forecast Hub were due~(black dashed line), human judgment predictions excluded in formal analysis~(dark blue), for what week each forecast was made~(red dash line), and the reported number of weekly incident COVID-19 cases at the US national level~(black solid line).
    (B.)~Forecasts of weekly incident cases submitted to the COVID-19 Forecast Hub~(orange) were formatted as seven quantiles, and we similarly formatted human judgment predictions from Metaculus~(blue) and Good Judgment Open~(red). 
    (C.)~Forecasts of weekly incident deaths submitted to the COVID-19 Forecast Hub were formatted as twenty three quantiles and we formatted human judgment predictions the same.
    We collected more than 3,000 original and revised  human judgement predictions of incident cases and deaths of the spread of SARS-CoV-2 and burden of COVID-19 in the US.
    \label{fig.experimentAndDataCollection}}
\end{figure}

The number of weeks between when a forecast was generated~(the forecast date) and the week when the truth would be determined~(the target end date) was 2 weeks for January, February, March, and April, and 3 weeks for May and June.
There were more than one forecast date we could have chosen between the start and close date of each survey.
We decided to chose the earliest forecast date that was the same as the COVID-19 forecast date~(Fig.~\ref{fig.experimentAndDataCollection}A.). 

Analyses below focus on predictions of incident cases which were formatted as 7 quantiles: 0.025, 0.100, 0.250, 0.500, 0.750, 0.900, 0.975~(Fig.~\ref{fig.experimentAndDataCollection}B.) and incident deaths which were formatted as 23 quantiles: 0.01, 0.025, quantiles from 0.05 to 0.95 in increments of 0.05, 0.975, and 0.99 at the national level~(Fig.~\ref{fig.experimentAndDataCollection}C.).
These 12 predictions were made by both human judgment and computational models at overlapping times.

\subsection{Ensemble and individual performance}

\begin{figure}[ht!]
    \centering
    \includegraphics{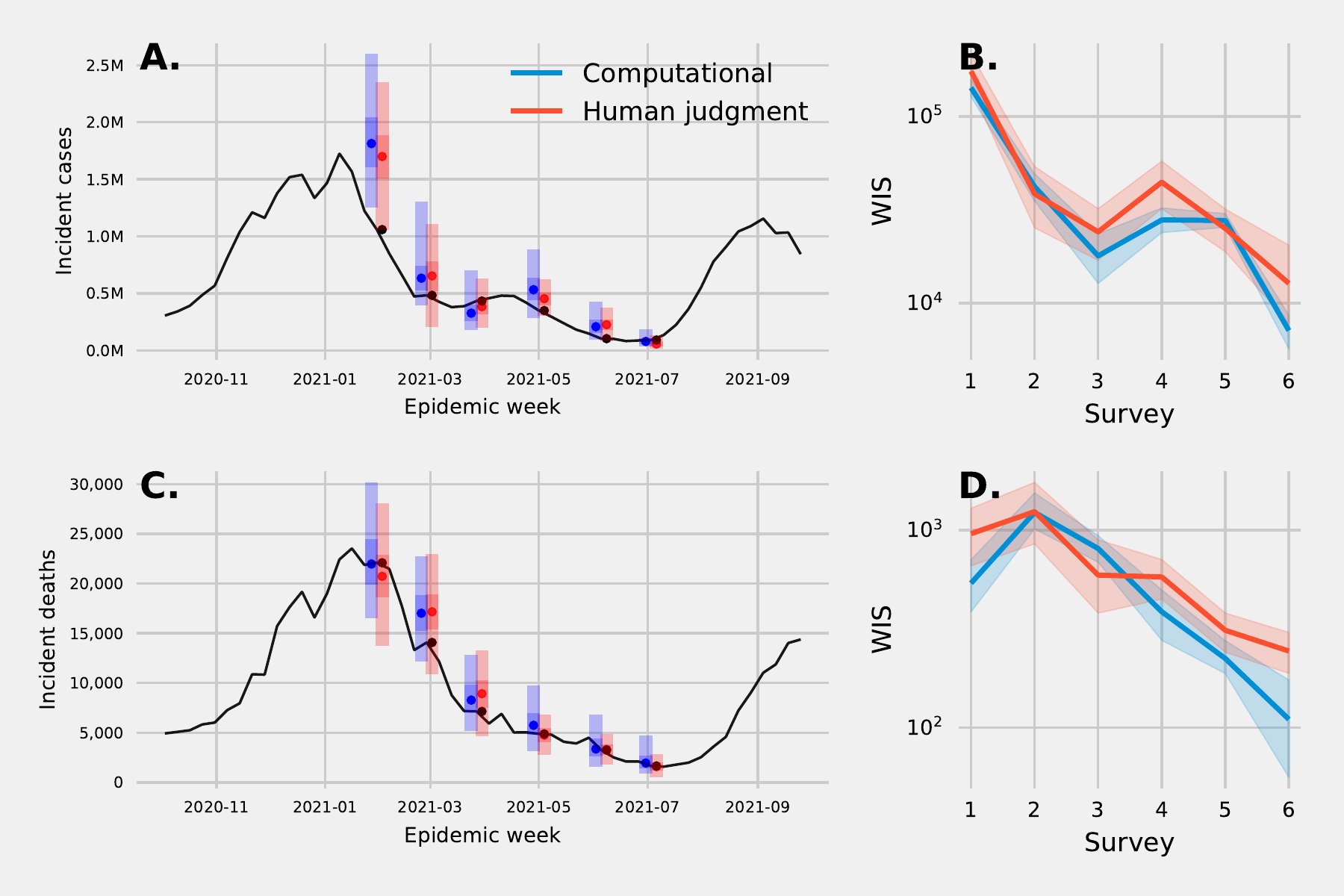}
    \caption{(A.)~Forecasts of weekly incident cases at the national level by an ensemble of computational models~(blue) and ensemble of human judgement~(red). 
    The dot represents the median forecast and the shaded bars represent the 25th and 75th, and the 2.5th and 97.5th prediction intervals.
    (B.)~A mean and 95\% confidence interval of the weighted interval score~(WIS) for forecasts of incident cases made by individual computational and human judgement models. 
    (C.)~Forecasts of weekly incident deaths and forecasts from computational models and human judgement.
    (D.)~Mean and 95\% confidence intervals of the WIS for individual predictions of incident deaths.
    Though individual human judgement forecasts tend to perform worse than computational models, a human judgement ensemble performed similar to an ensemble of computational models for predictions of both cases and deaths over a 6 month period.~\label{fig.epicurve}}
\end{figure}

An ensemble of human judgment models made similar two and three week ahead predictions of weekly incident cases and deaths at the national level when compared to a computational ensemble~(Fig.~\ref{fig.epicurve}A. and C.) despite individual human judgement predictions performing slightly worse on average~(Fig.~\ref{fig.epicurve}B. and D.).

The median prediction of incident cases was closer to the truth on more occasions for human judgement compared to computational models~(Fig.~\ref{fig.epicurve}A.).
Human judgement and computational ensembles both overestimated incident cases in late January and to a lesser extent they overestimated the number of cases in February and May.
For all six surveys the median prediction for  computational models and human judgment were both larger or smaller than the truth.
Though the human judgement ensemble median prediction is at times closer to the truth than the compuational ensemble, the mean WIS score for individual predictions across all but one survey is smaller for computational models than for human judgement~(Fig.~\ref{fig.epicurve}B.).

The median prediction of incident deaths was at times closer to the truth for computational models and at other times closer for a human judgement ensemble~(Fig.~\ref{fig.epicurve}C.).
January to May median predictions for computational models assumed a shallower decline in the number of deaths when compared to human judgement predictions for which the median prediction remained higher than the truth for predictions in January, February, and March, and then smaller than the truth in April. 
For one time point, the week beginning April 25th and ending May 1st, the median prediction from a computational ensemble was above the truth and the median predictions for human judgement was below the truth.
Again, the mean WIS score for individual computational models is smaller when compared to human judgement, though the median prediction is at times closer to the truth for computational models and at times closer for human judgement~(Fig.~\ref{fig.epicurve}D.)


\subsection{Pattern of missing forecasts for computational and human judgment models}

\begin{figure}[ht!]
    \centering
    \includegraphics{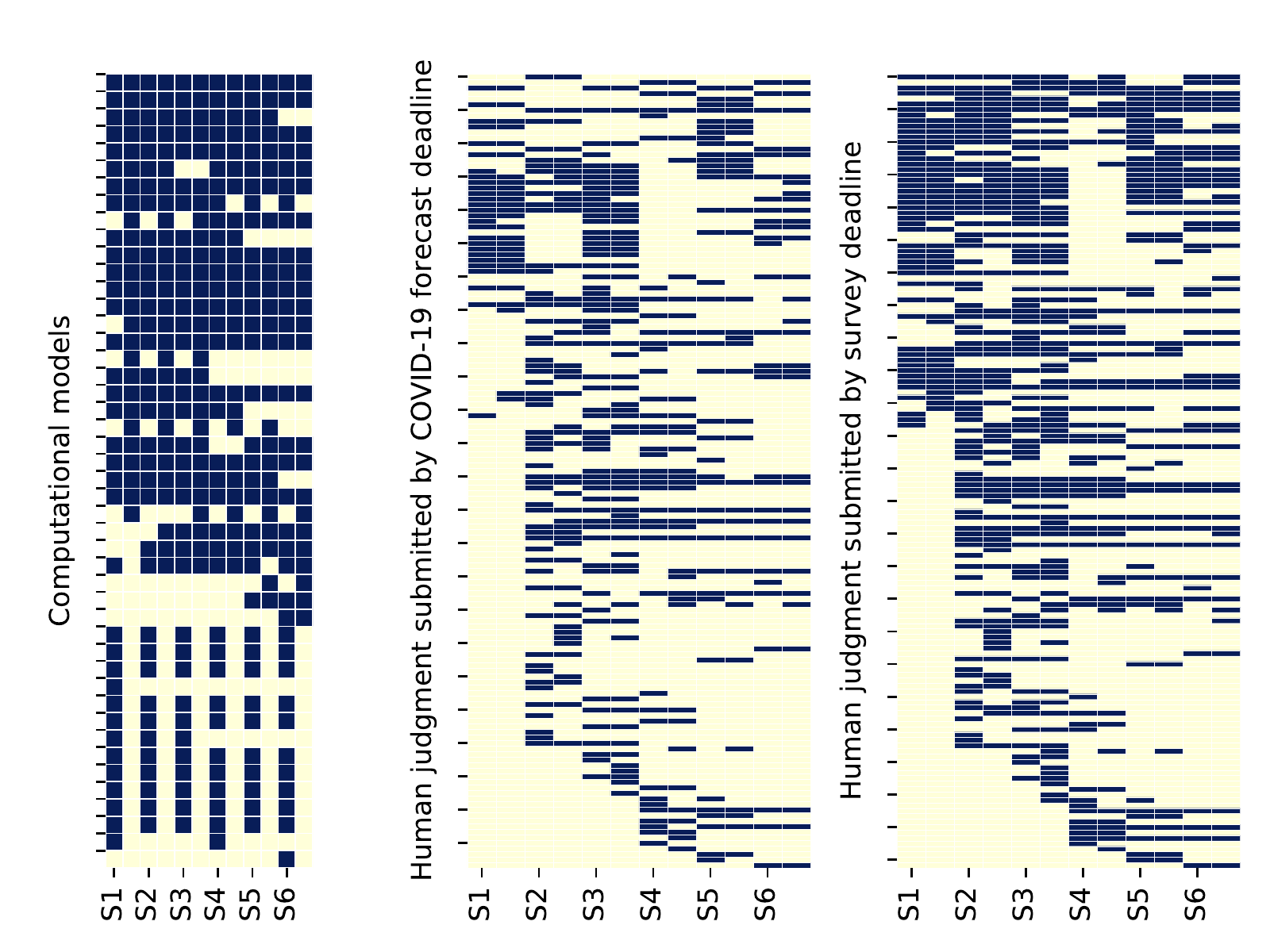}
    \caption{Submitted and missing forecasts made by (A.)~computational forecasts, (B.)~human judgment forecasts submitted before the COVID-19 deadline, and (C.)~human judgment forecasts submitted by the survey deadline.
    Forecasts that were submitted are shown in blue and forecasts not submitted (missing) are shown in yellow.
    Rows represent a single model and columns are broken into six pairs---the left column (with the tick mark) corresponds to submissions of incident cases and the second column in the pair corresponds to submissions of incident deaths---which represent the six surveys from January 2021 to June 2021.
    The high proportion of missing forecasts made by human judgement models presents a methodological challenge when building a chimeric ensemble. 
    \label{fig.patternofmissing}}
\end{figure}

The mean proportion of missing forecasts per model is higher for human judgment forecasts that submitted predictions at or before the forecast date set by the COVID-19 Forecast Hub~(71\%) versus computational models~(34\%): t-stat = 8.92, pvalue $<$0.001~(Fig.~\ref{fig.patternofmissing}).
The mean proportion of missing human judgment forecasts per model made by the survey deadline was smaller~(66\%) than was made by the COVID-19 Forecast Hub deadline~(71\%).



The proportion of surveys submitted by human judgment models compared to computational models that included both a prediction for cases and deaths was 23\% vs 49\%, that included a prediction for either cases or (exclusive) deaths is 11\% vs 33\%, and that did not submit both cases and deaths was 65\% vs 17\%.
    

\subsection{Comparison of a chimeric and computational ensemble and the impact of imputation}

\begin{figure}[ht!]
    \centering
    \includegraphics{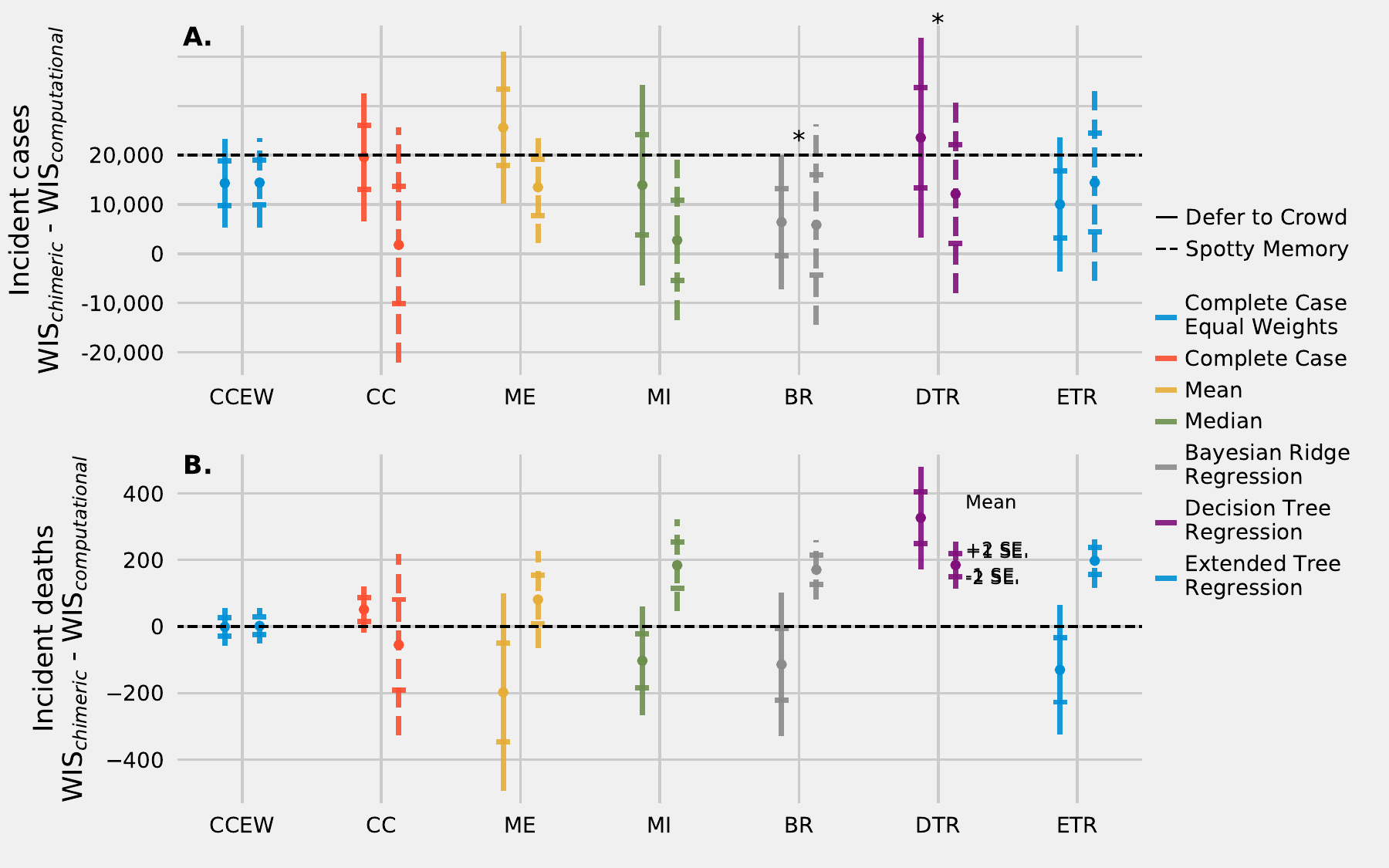}
    \caption{ Mean difference in WIS for incident cases~(A.) and deaths~(B.) at the US national level between a chimeric ensemble and a computational ensemble paired across six different surveys from Jan 2021 to June 2021 for two strategies to impute missing values ("spotty memory" and "defer to the crowd") and, within each strategy, 5 different techniques to impute missing forecasts. Two asterisks denote a pvalue smaller than 0.05 from a one-sided paired t-test. A chimeric ensemble---a combination of computational and human judgment models---improves WIS scores when the target is cases but weakens or maintains similar WIS scores when the target is deaths. There are negligible differences in mean WIS between a "defer to the crowd" and "spotty memory" imputation strategy for prediction of cases and a defer to the crowd approach appears to improve predictions compared to a spotty memory approach for predictions of incident deaths. Bayesian Ridge Regression~(BR) and Median imputation (MI) are promising strategies to impute missing forecasts for incident cases.\label{fig.headtohead}}
\end{figure}

A chimeric ensemble improved predictions of incident cases compared to an computational model only ensemble.
The mean WIS score assigned to predictions of incident cases for a chimeric ensemble minus the WIS score for a computational model paired by survey was negative~(i.e. was improved) when using specific imputation techniques and strategies, and for the complete case~(Fig.~\ref{fig.headtohead}~A.).
Imputing forecasts with a decision tree regression~(DTR) and "defer to the crowd" strategy had the smallest mean paired WIS score (mean: -6,111; t-stat = -2.13) and a paired t-test suggests this result is significant (pvalue $<$ 0.043). 
Imputing missing predictions using a bayesian ridge regression~(BR) also performed well. 
A complete case equally weighted~(CCEW in Fig.~\ref{fig.headtohead}) chimeric ensemble reported similar predictive performance compared to an equally weighted computational ensemble using a "defer to the crowd" approach~(mean, paired WIS: -2,760; t-stat=-1.21; pvalue=0.14) and when using a "spotty memory" strategy~(mean, paired WIS: -2,747; t-stat = -1.21; pvalue=0.14).  
Weighting a combination of computational and human judgment models, coupled with an imputation strategy, may better predict incident cases at the US national level compared to a computational model only ensemble.

In contrast to incident cases, the paired mean WIS score for incident deaths was positive~(i.e. performed worse) or close to zero for the majority of imputation strategies, the complete case dataset, and a complete case data set where equal weights are assigned to all models~(Fig.~\ref{fig.headtohead}~B.).
A chimeric ensemble may not improve predictions of incident deaths compared to an ensemble of computational models alone.


\subsection{Performance based vs equal weighting}

\begin{figure}[ht!]
    \centering
    \includegraphics{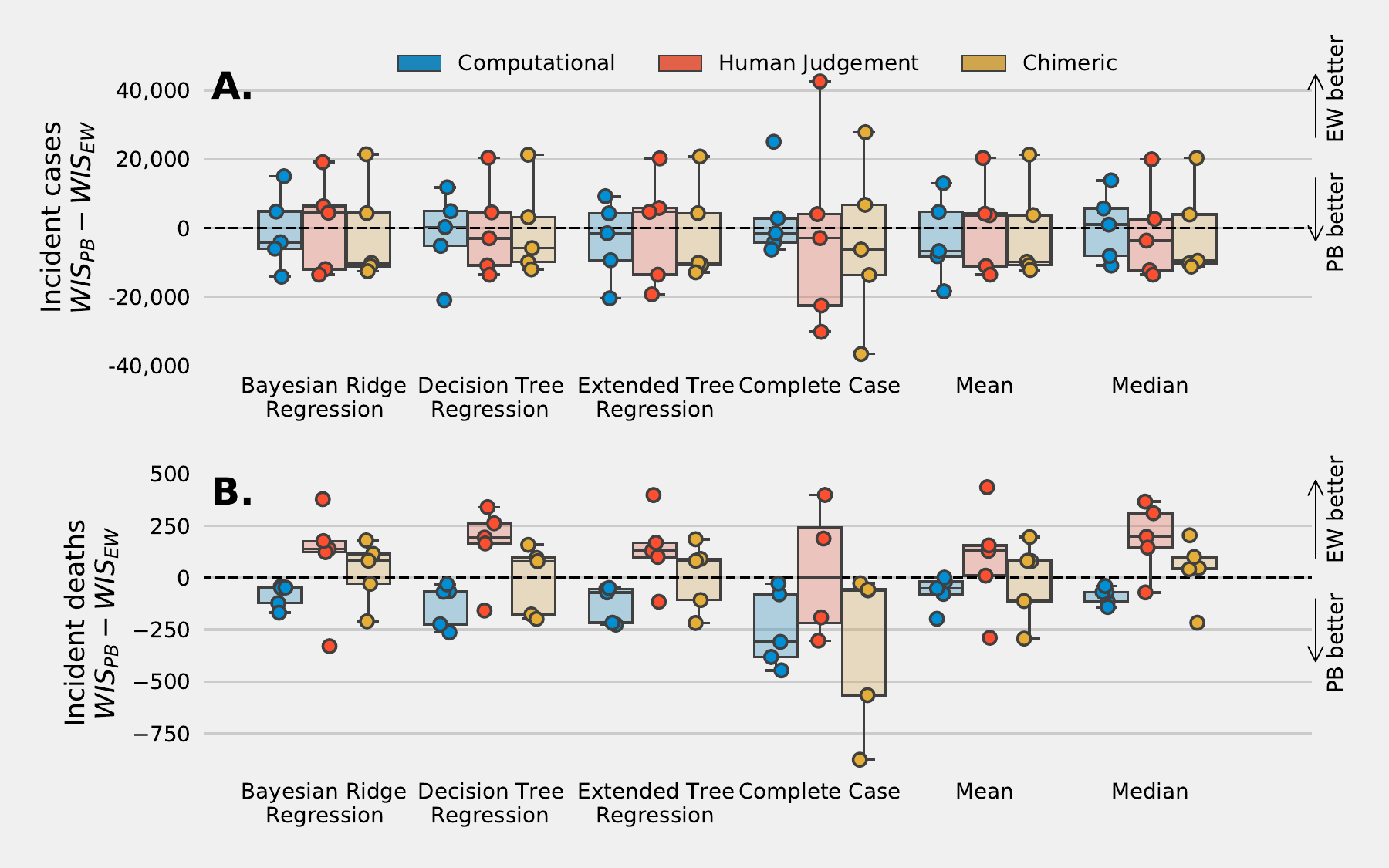}
    \caption{Median, 25th and 75th, and interquartile ranges for the difference between WIS scores when fitting a performance based ensemble~(PB) and equally weighted ensemble~(EW) paired by survey for three different ensembles: an ensemble that includes only computational models~(blue), only human judgment~(red), and a chimeric ensemble that includes both computational and human judgement models~(gold). 
    A "spotty memory" strategy was used along with five imputation techniques for training. Ensemble predictions are stratified by (A.)~ incident cases and (B.)~deaths.
    For the majority of imputation techniques used for predictions of incident cases, training a performance based ensemble shows similar results for a chimeric, computational, and human judgement ensemble.  
    For deaths, performance based training improves predictions of a computational ensemble, shows little improvement to a chimeric ensemble, and weakens predictions of a human judgment ensemble.
    \label{fig.perfvsequal_spotty}}
\end{figure}

A performance based ensemble~(PB) compared to assigning to all models equal weights~(EW) decreases median WIS score for predictions of US national incident deaths when considering a computational ensemble, but not a chimeric or human judgement ensemble using a spotty memory imputation strategy. 
For all three ensembles WIS scores for predictions of cases show similar performance weights compared to equal weights ~(Fig.~\ref{fig.perfvsequal_spotty}).

\begin{figure}[ht!]
    \centering
    \includegraphics{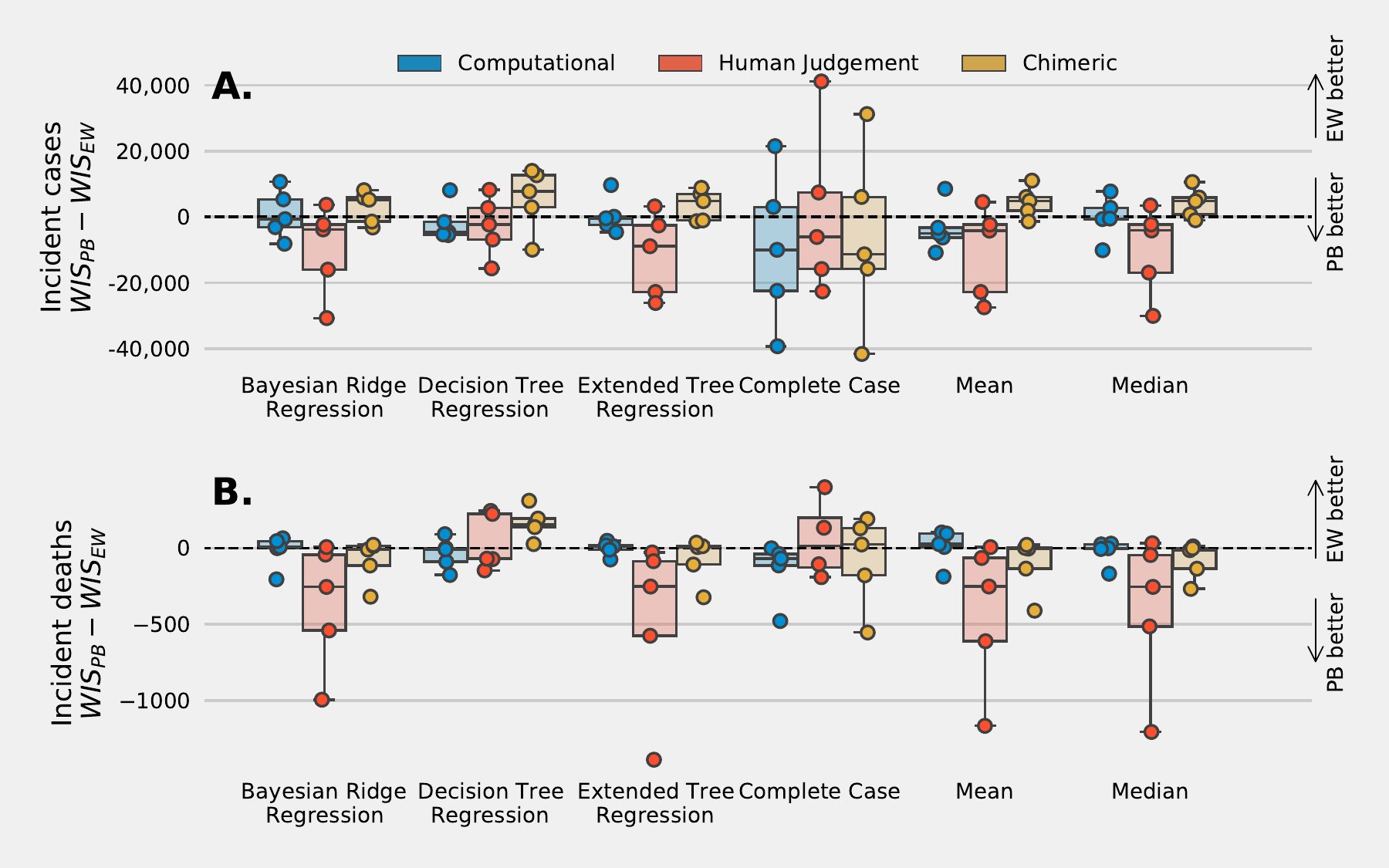}
    \caption{Median, 25th and 75th, and interquartile ranges for the difference between WIS scores when fitting a performance based ensemble~(PB) and equally weighted ensemble~(EW) paired by survey for three different ensembles: an ensemble that includes only computational models~(blue), only human judgment~(red), and a chimeric ensemble that includes both computational and human judgement models~(gold). 
    A "defer to the crowd" strategy was used along with five imputation techniques for training. Ensemble predictions are stratified by (A.)~ incident cases and (B.)~deaths.
    For the majority of imputation techniques used for predictions of incident cases, training a performance based ensemble improves the WIS score of a human judgement ensemble and weakens the performance of a computational and chimeric ensemble.  
    For deaths, performance based training improves predictions of a a chimeric and human judgement ensemble, but for some imputation techniques weakens predictions of a computational ensemble. An algorithm that assigns different weights based on past performance, coupled with a "defer to the crowd" imputation strategy, may improve predictive performance of a chimeric ensemble.
    \label{fig.perfvsequal_defer}}
\end{figure}

For predictions of incident cases with a spotty memory imputation strategy~(Fig.~\ref{fig.perfvsequal_spotty}A.), the median across all imputation techniques is negative, and  the 25th to 75th percentiles include zero, indicating that performance based weighting is similar for predictions of incident cases. 
A defer to the crowd approach plus performance weighting improves predictions for a human judgement ensemble and for a computational ensemble, but weakens predictive performance for a chimeric ensemble~(Fig.~\ref{fig.perfvsequal_defer}A.).

For predictions of incident deaths, a performance based ensemble plus spotty memory approach improves WIS scores for a computational ensemble, shows similar performance for a chimeric ensemble, and weakens performance of a human judgement ensemble~(Fig.~\ref{fig.perfvsequal_spotty}B.).
A defer to the crowd approach plus performance weights improves human judgement and chimeric ensemble performance and weakens the performance of a computational ensemble~(Fig.~\ref{fig.perfvsequal_defer}B.)
A complete case strategy plus performance weights shows similar WIS scores when using a human judgement and chimeric ensemble and improves predictions when using a computational and chimeric ensemble.
The interquartile range for $\text{WIS}_{\text{PB}}-\text{WIS}_{\text{EW}}$ 
is above or covers zero for most chimeric and human judgment ensembles and is below zero for a computational ensemble when using a complete case approach.

\clearpage
\section{Discussion}

We presented a first effort to combine direct probabilistic predictions of the spread and burden of an infectious agent generated by both computational models and human judgement.

A chimeric ensemble---a combination of forecasts generated by computational models and human judgment models---is capable of producing predictions that outperform an ensemble of computational models only.
Though a chimeric ensemble has the potential to outperform a computational ensemble this is not always the case. 
Chimeric ensemble modeling is still in early stages and the reader should consider this work hypothesis generating.

There are several challenges to overcome when adding human judgment predictions.

Human judgment data must first be collected before predictions can be combined to produce a forecast.
Data collection requires a team to pose questions to an audience of forecasters.
Questions should be written as clear and concise as possible, to minimize bias, and written so that the forecaster understands how the truth will be determined~(often called the resolution criteria).  
After questions are drafted they must be submitted to a prediction platform.
A prediction platform should allow forecasters to easily view the question and resolution criteria, and allow the forecaster to submit their prediction with minimal effort. 
An immense amount of time and effort is needed to draft questions, and build and host a prediction platform.  
Organizing computational modeling efforts too requires an immense amount of effort to build~\cite{reich2019collaborative,ray2020ensemble,mcgowan2019collaborative}. However, the time needed to host computational efforts and answer questions throughout the prediction period may be less burdensome than with a human judgement platform.

After data collection there continue to be challenges with human judgment predictions.
In our opinion, the most pressing issue is missing forecasts. 
Compared to computational models, we found that human forecasters have a much higher rate of missing forecast submissions, and if one wishes to use only models that submitted all forecasts~(a complete case approach) it may not be feasible to include human judgment. 
Instead, an imputation strategy should be used to account for missing human judgment forecasts.
Here we proposed two potential strategies to account for missing forecasts: a "defer to the crowd" and "spotty memory" approach, and we found that both methods resulted in similar predictive performance of incident cases and deaths for most imputation functions, though the "defer to the crowd" strategy may produce more accurate predictions of cases when using a bayesian regression function to impute missing values and a spotty memory approach produced the most accurate forecasts when using median imputation.
Both methods were able to incorporate more human judgment models in an ensemble than a complete case analysis. 
That said, the chimeric ensemble using a complete case approach with equal weights---the most natural approach--- showed improved performance compared to a computational ensemble and is one of the best pieces of evidence that adding human judgement can improve forecasts of an infectious agent.

The need to couple ensemble modeling with an imputation strategy is not unique to chimeric forecasts, but we feel the proportion of missing forecasts is unique~\cite{mccandless2011effects}.
Because the imputation strategies often fill in missing forecasts for a specific target with similar quantile values, one could consider the imputation approach we took to be a type of regularization and in past literature regularization was found to improve computational and human judgement ensembles~\cite{mcandrew2021adaptively,merkle2020beating}.

Whether to use a performance based or equal weighting for a chimeric ensemble is still unclear.
A performance based chimeric ensemble compared to an equally weighted ensemble showed improved performance for some surveys and weakened performance for other surveys using a spotty memory approach~(Supplemental Fig.~\ref{fig.pairedOverSurveys_spotty}), and showed improved performance as additional data was collected for a defer to the crowd approach coupled with a chimeric ensemble when predicting cases~(Supplemental Fig. \ref{fig.pairedOverSurveys_defer}).
A challenge when ensemble modeling, in addition to choosing an algorithm to assign different weights to models, is to know in advance whether or not differential weighting will improve predictive performance and whether or not human judgement will improve or weaken predictive performance. 
Some factors that may help determine if differential weighting is useful or if human judgement should be included could be the difference in predicted median between a computational ensemble and human judgement ensemble, or potentially the difference in uncertainty in predictions.
More work should focus on a three step approach to ensemble modeling: (i) predicting whether human judgement will improve predictive performance, (ii) predicting if differential weighting would benefit a set of models, and (iii) then either choosing equal weights or differential weights.

A chimeric and human judgement ensemble's ability to improve predictions of incident cases is consistent with past work studying predictions of exclusively human judgment~\cite{bosse2021comparing}.
Computational models often make more accurate predictions of deaths because they incorporate into their models reported cases, a signal for upcoming deaths.
We are not sure whether or not humans considered the time series of incident cases when submitting predictions of deaths.
Questions presented to forecasters did not suggest that cases could be a strong signal to consider when building a forecast for deaths.
The question of how forecasters use time series information could lead to a controlled experiment to test human judgment's ability to predict one time series by using a second, correlated time series. 
Previous literature suggests humans may make strong predictions that are short term, when there exists linear correlations between two concepts, and focus on information that most differed from their expectations~\cite{lagnado2006insight,hammond1965cognitive,spicer2021theory}.
But to the best of our knowledge no work has been done in the area of multi-cue probability theory and judgemental forecasting of time series by providing a second correlated time series.

Because the effort a human can spend on prediction is finite, and because of the above results that show human judgement improves predictions of cases the most, we recommend asking crowds to predict cases or similar targets that are strongly correlated to others~(such as incident deaths) which may (i) improve predictions of cases and (ii) improve predictions of deaths if these human judgement predictions were used as input to a computational forecasting model.

This work has several limitations. 
We only evaluated twelve targets in common with the COVID-19 Forecast hub and so the results above should be considered exploratory rather than confirmatory.
The limited number of targets brings up the broader limitation that human judgement cannot be applied to a large number of targets, locations, and forecast horizons like computational models.
The ensemble model we chose to optimize average WIS was deterministic, made no attempt to regularize weights assigned to models, and is just one type of method to aggregate computational and human judgement models. 
The number of human judgement participants, while excellent, was still a limitation at times.
The empirical nature of this work, versus a controlled laboratory experiment, as well makes it difficult to draw strong conclusions about the performance of human judgement, computational models, and their combined performance.

In the future we plan to focus on methodology: (i) by building more advanced ensemble algorithms to combine computational and human judgement models, (ii) methods to determine for which targets human judgement is needed and which targets it is not needed, and (iii) imputation procedures that take into account the uncertainty when filling in missing forecasts;  data collection: (i) by proposing strategies to reduce the number of missing human judgement forecasts; explore the limits of human judgement: (i) by testing  to what degree humans can use one time series to predict another, (ii) how humans construct mental models and generate predictions, and (iii) what additional information can human judgement provide that is supportive of public health efforts. 

We envision a chimeric ensemble as a flexible model that can manage and combine predictions throughout the evolution of an infectious agent and as a supportive tool for public health.
A chimeric ensemble can begin to support primary and secondary preventive measures by relying on fast acting human judgment to forecast targets while data is collected and computational models are trained.
Once computational models begin to forecast, a chimeric ensemble can integrate these forecasts with no down time.
As computational models become accurate for specific targets then human judgement can be used to predict noisier targets which can be included in this type of ensemble. 

\section{Acknowledgements}

This research was supported through the MIDAS Coordination Center (MIDASNI2020- 1) by a grant from the National Institute of General Medical Science (3U24GM132013-02S2). 
We wish to thank Nikos Bosse, Estee Cramer, Chris Karvetski for useful comments that improved the quality of this work, and to those at the COVID-19 Consortium Colloquium Speaker Series hosted by the UT COVID-19 Modeling Consortium at the University of Texas, Austin who offered great insights incorporated in this work. We wish to thank Phillip Rescober for data science support from Good Judgement Inc. Finally, we wish to thank all of the individual forecasters who contributed their time and energy to generate predictions about the trajectory of COVID-19.

\section{Tables}

\begin{table}[ht!]
    \centering
    \begin{tabular}{lp{5cm}p{5cm}}
        \hline
        Imputation technique & \parbox[c]{5cm}{G} & \parbox[c]{5cm}{Summary}  \\
        \hline
        Mean   & \parbox[c]{5cm}{$I^{-1}\sum_{i}{q_{i,k}}$} & \parbox[l]{5cm}{Take the mean of all present quantiles where the set $I$ is an index for present forecasts} \vspace{4mm}\\  
        Median & \parbox[l]{5cm}{$\min_{x} \l\{ F(x)-1/2 \r\}$}  & \parbox[c]{5cm}{Take the median of all present quantiles where $F$ is the empirical cdf over all $I$ quantiles} \vspace{4mm}\\
        Bayesian Ridge regression &  $ \mathbb{E}(X) \text{ where } X\sim \mathcal{N}(Q_{-k}\beta,\sigma^{2})$\\
                                  &   $\beta \sim \mathcal{N}(0,\lambda^{-1} I) \;\sigma^{2} \sim \Gamma(\alpha,\gamma)$
                                & \parbox[c]{5cm}{The matrix $Q_{-k}$ has two columns: a column of ones and a second column of quantiles from present forecasts.} \vspace{4mm}\\
        Decision Tree regression & - & The missing quantile value is imputed by the mean of quantiles in the same partition. \vspace{4mm}\\
        Extremely Randomized Trees & - & Multiple decision trees $(D_{i})$ are fit to random subsets of quantiles and the missing forecast is imputed as the average over $D_{i}$.\\ 
        \hline
    \end{tabular}
    \caption{Five procedures were chosen to impute missing forecasts. Mean and median imputation only use information about a single quantile to impute missing forecasts, while the three regression approaches use all the quantiles from all present forecasts to impute missing forecasts. \label{tab.impute}}
\end{table}

\clearpage
\section{Figures}

\clearpage

\appendix

\clearpage
\section{Questions and Resolution Criteria}
\label{tab.quesres}

\subsection{Survey 01 (Open on 2021-01-06, Close on 2021-01-16)}
\begin{itemize}
    \item Question 1
        \begin{itemize}
            \item Question:  What will be the number of new incident U.S. adult and pediatric previous day admissions to the hospital with confirmed COVID-19 for the week beginning on 2021-01-24 and ending on 2021-01-30 (inclusive)?
            \item Resolution Criteria:This question will resolve as the total number of adult plus pediatric previous day admissions with confirmed COVID-19 as recorded in the Department of Health and Human Service’s report of COVID-19 reported patient impact and hospital capacity. The total previous day admissions is computed using two variables in this report: previous\_day\_admission\_adult\_covid\_confirmed and previous\_day\_admission\_pediatric\_covid\_confirmed. This report, and the resolution criteria, includes data on all 50 US states, Washington DC, Puerto Rico, and the US Virgin Islands (53 states and territories). The report will be accessed one week after the end of the month (2021-02-06).
            \item Range:[0-240K]
            \item Question URL: \url{https://pandemic.metaculus.com/questions/6160/new-us-hospital-admissions-24-30-january/}
        \end{itemize}
    \item Question 2
        \begin{itemize}
            \item Question: What will be the total number of new incident deaths due to COVID-19 in the US for the week beginning on 2021-01-24 and ending on 2021-01-30 (inclusive)?
            \item Resolution Criteria: This question will resolve as the number of new deaths due to confirmed COVID-19 for the week beginning on 2021-01-24 and ending on 2021-01-30 (inclusive) as recorded in the Johns Hopkins University (JHU) CSSE Github data repository. This file records the daily number of deaths by county. From this file deaths are summed across all counties and aggregated by week to generate the number of new deaths per week. The report will be accessed one week after the end of the month (2021-02-06).
            \item Range:[0-40k]
            \item Question URL: 
            \url{https://pandemic.metaculus.com/questions/6161/new-us-covid-deaths-24-30-january/}
        \end{itemize}
    \item Question 3
        \begin{itemize}
            \item Question: What will be the number of new incident confirmed positive COVID-19 cases in the US beginning on 2021-01-24 and ending on 2021-01-30 (inclusive)?
            \item Resolution Criteria: This question will resolve as the number of new confirmed cases beginning on 2021-01-24 and ending on 2021-01-30 (inclusive) recorded in the Johns Hopkins University (JHU) CSSE Github data repository. This file records the daily number of cases by county. From this file cases are summed across all counties and aggregated by week to generate the number of new cases per week. The report will be accessed one week after the end of the month (2021-02-06).
            \item Range:[0-3M]
            \item Question URL:
            \url{https://pandemic.metaculus.com/questions/6162/new-us-covid-cases-24-30-january/}
        \end{itemize}
    \item Question 4
        \begin{itemize}
            \item Question: What factor should the median 4-week-ahead COVIDhub Ensemble forecast of national incident deaths made on 4 Jan(a forecast for the 24-30 Jan week) be multiplied by so that it equals the reported number of new US incident deaths?
            \item Resolution Criteria: This question will resolve as the factor that the The COVID-19 Forecast Hub's “COVIDhub” Ensemble median forecast of the US national number of incident deaths 4 weeks into the future should be multiplied by (the reported number of incident deaths divided by the forecasted median) to equal the reported number of US national incident deaths as reported by the Johns Hopkins University (JHU) CSSE Github data repository.
            \item Range:[0-3]
            \item Question URL
            \url{https://pandemic.metaculus.com/questions/6163/factor-covidhub-forecast-to-be-multiplied-by/}
        \end{itemize}
    \item Question 5 
        \begin{itemize}
            \item Question: What will be the cumulative number of deaths due to COVID-19 on 2021-12-31 if less than 50\% of Americans initiate vaccination (1st dose received) with a COVID-19 vaccine by 2021-03-01?
            \item Resolution Criteria: The percent of the population that received a COVID-19 vaccine on or before 2021-03-01 will be computed by dividing the number of individuals who have initiated vaccine (1st dose taken) provided by the CDC COVID data tracker by the current US population which on 2021-01-04 was reported to be 330,782,991 and multiplying this fraction by 100. The CDC COVID data tracker that counts the number of individuals who have initialized vaccination will be accessed when data is available after and as close as possible to 2021-03-01.

            To resolve deaths, we will use the cumulative number of deaths due to confirmed COVID-19 as recorded in the Johns Hopkins University (JHU) CSSE Github data repository. This file records the daily number of deaths by county. From this file deaths are summed across all counties and aggregated to week to generate the number of new deaths per week. The report will be accessed one week after 2021-12-31.

            9 January edit: This question will resolve ambiguously if greater than or equal to 50\% of Americans are vaccinated by 2021-03-01.
            
            \item Range:[0-1M]
            \item Question URL:
            \url{https://pandemic.metaculus.com/questions/6165/number-of-us-covid-deaths-if-50-vaccinated/}
        \end{itemize}
    \item Question 6
        \begin{itemize}
            \item Question: What will be the cumulative number of deaths due to COVID-19 on 2021-12-31 if greater than or equal to 50\% of Americans initiate vaccination (1st dose received) with a COVID-19 vaccine by 2021-03-01?
            \item Resolution Criteria: The percent of the population that received a COVID-19 vaccine on or before 2021-03-01 will be computed by dividing the number of individuals who have initiated vaccine (1st dose taken) provided by the CDC COVID data tracker by the current US population which on 2021-01-04 was reported to be 330,782,991 and multiplying this fraction by 100. The CDC COVID data tracker that counts the number of individuals who have initialized vaccination will be accessed when data is available after and as close as possible to 2021-03-01.

            To resolve deaths, we will use the cumulative number of deaths due to confirmed COVID-19 as recorded in the Johns Hopkins University (JHU) CSSE Github data repository. This file records the daily number of deaths by county. From this file deaths are summed across all counties and aggregated to week to generate the number of new deaths per week. The report will be accessed one week after 2021-12-31.

            9 January edit: This question will resolve ambiguously if less than 50\% of Americans are vaccinated by 2021-03-01.
            \item Range:[0-1M]
            \item Question URL:
            \url{https://pandemic.metaculus.com/questions/6164/number-of-covid-deaths-if-50-vaccinated/}
        \end{itemize}
    \item Question 7
        \begin{itemize}
            \item Question: What will be the percent of B.1.1.7 among all S gene dropout SARS-CoV-2 samples submitted for genomic sequencing surveillance in the US between 2021-02-01 and 2021-02-15?
            \item Resolution Criteria: This NextStrain page on S:N501Y as a proportion of overall S gene dropout samples in the US will be consulted for resolution. To access this data, scroll to the bottom of the page and download “selected metadata.” In this Excel sheet, please consult the “Pangolin Lineage” column to find B.1.1.7 samples. Samples that have a “collection data” date between 2021-02-01 and 2021-02-15 (inclusive) will be considered. We will compute the proportion by dividing the number of B.1.1.7 samples by the total number of samples (which are all 69del samples) in the spreadsheet. NextStrain will be accessed in the last week of February (the week ending on 2021-02-27) for resolution.
            \item Range:[0-100]
            \item Question URL:
            \url{https://pandemic.metaculus.com/questions/6166/-b117-among-all-s-gene-dropout-samples/}
        \end{itemize}
\end{itemize}

\subsection{Survey 02 (Open on 2021-02-03, Close on 2021-02-15)}
\begin{itemize}
    \item Question 1
        \begin{itemize}
            \item Question: What will be the number of new incident U.S. adult and pediatric admissions to the hospital with confirmed COVID-19 for the week beginning on 2021-02-21 and ending on 2021-02-27 (inclusive)?
            \item Resolution Criteria: This question will resolve as the total number of adult plus pediatric previous day admissions with confirmed COVID-19 as recorded in the Department of Health and Human Service’s report of COVID-19 reported patient impact and hospital capacity for the dates from 2021-02-22 to 2021-02-28, corresponding to the number of hospitalizations from 2021-02-21 to 2021-02-27. Daily updates are provided by the Department of Health and Human Services. The total previous day admissions is computed using two variables in this report: previous\_day\_admission\_adult\_covid\_confirmed and previous\_day\_admission\_pediatric\_covid\_confirmed and stored in Lehigh University's Computational Uncertainty Lab Github data repository. This report, and the resolution criteria, includes data on all 50 US states, Washington DC, Puerto Rico, and the US Virgin Islands (53 states and territories). The report will be accessed no sooner than 2021-03-06.
            \item Range:[0-200K]
            \item Question URL Metaculus: \url{https://pandemic.metaculus.com/questions/6468/new-us-covid-hospital-admissions-21-27-feb/}
            \item Question URL GJO: \url{https://www.gjopen.com/questions/1921}
        \end{itemize}
    \item Question 2
        \begin{itemize}
            \item What will be the total number of new incident deaths due to COVID-19 in the US for the week beginning on 2021-02-21 and ending on 2021-02-27 (inclusive)?
            \item Resolution Criteria: This question will resolve as the number of new deaths due to confirmed COVID-19 for the week beginning on 2021-02-21 and ending on 2021-02-27 (inclusive) as recorded in the Johns Hopkins University (JHU) CSSE Github data repository. This file records the daily number of deaths by county. From this file deaths are summed across all counties and aggregated by week to generate the number of new deaths per week. The number of deaths for the week beginning on 2021-02-21 will be computed by adding the number of new deaths from the 2021-02-21 up to, and including, 2021-02-27. The report will be accessed no sooner than (2021-03-06).
            \item Range:[0-30k]
            \item Question URL Metaculus: 
              \url{https://pandemic.metaculus.com/questions/6466/new-us-covid-deaths-21-27-february/}
            \item Question URL GJO: \url{https://www.gjopen.com/questions/1922}
        \end{itemize}
    \item Question 3
        \begin{itemize}
            \item Question: What will be the total number of new incident confirmed positive COVID-19 cases in the US beginning on 2021-02-21 and ending on 2021-02-27 (inclusive)?
            \item Resolution Criteria: This question will resolve as the number of new confirmed cases beginning on 2021-02-21 and ending on 2021-02-27 (inclusive) recorded in the Johns Hopkins University (JHU) CSSE Github data repository. This file records the daily number of cases by county. From this file cases are summed across all counties and aggregated by week to generate the number of new cases per week. The report will be accessed no sooner than 2021-03-06.
            \item Range:[0-3M]
            \item Question URL Metaculus:
            \url{https://pandemic.metaculus.com/questions/6469/new-us-covid-cases-21-27-february/}
            \item Question URL GJO: \url{https://www.gjopen.com/questions/1923}
        \end{itemize}
    \item Question 4
        \begin{itemize}
            \item Question: What will be the cumulative number of people who have received one or more doses of a COVID-19 vaccine in the U.S. on 2021-02-28?
            \item Resolution Criteria: This question will resolve as the cumulative number of people who receive one or more doses of a COVID-19 vaccine on 2021-02-28 as recorded by the Centers for Disease Control COVID-19 Data tracker in the column "Number of People Receiving 1 or More Doses." The dashboard is updated daily by 8pm ET and will be accessed on 2021-02-28 at approximately 10:00pm ET.
            \item Range:[27M-85M]
            \item Question URL Metaculus:
              \url{https://pandemic.metaculus.com/questions/6472/cumulative-us-vaccinations-28-february/}
            \item Question URL GJO: \url{https://www.gjopen.com/questions/1924}
        \end{itemize}
    \item Question 5 
        \begin{itemize}
            \item Question: How many variants of concern will be monitored by the US CDC as of 2021–03-07?
            \item Resolution Criteria: This question will resolve as the number of variants of concern at the following link: “US COVID-19 Cases Caused by Variants” page as of Sunday, 2021–03-07. For example, as of 2021–02-02 this page shows that there are three variants: B.1.1.7, B.1.351, and P.1. This page is updated on Sundays, Tuesdays, and Thursdays by 7pm ET and will be accessed at approximately 10pm ET on 2021–03-07 (a Sunday).
            \item Range:[0-8]
            \item Question URL:
            \url{https://pandemic.metaculus.com/questions/6474/-variants-monitored-by-cdc-on-7-march/}
        \end{itemize}
    \item Question 6
        \begin{itemize}
            \item Question: What will be the percent of S:N501 sequences in the U.S. among all positive SARS-CoV-2 samples submitted to the GISAID database of genetic sequences between 2021-03-01 and 2021-03-07 (inclusive)?
            \item Resolution Criteria: This question will resolve as the percentage of US S:N501 sequences among all positive SARS-CoV-2 samples submitted for genomic sequencing to the GISAID database between 2021-03-01 and 2021-03-07 (inclusive), as displayed on the "Distribution of S:N501 per country" plot on following website: https://covariants.org/variants/S.N501. This website pulls data from GISAID and makes it publicly accessible. This percentage will be accessed no sooner than 2021-03-15.
            \item Range:[0-100]
            \item Question URL:
            \url{https://pandemic.metaculus.com/questions/6473/-sn501-in-us-for-week-of-1-march/}
        \end{itemize}
    \item Question 7
        \begin{itemize}
            \item Question: What will be the percent of S:N501 sequences in the U.S. among all positive SARS-CoV-2 samples submitted to the GISAID database of genetic sequences between 2021-03-29 and 2021-04-04 (inclusive)?
            \item Resolution Criteria: This question will resolve as the percentage of US S:N501 sequences among all positive SARS-CoV-2 samples submitted for genomic sequencing to the GISAID database between 2021-03-29 and 2021-04-04 (inclusive), as displayed on the "Distribution of S:N501 per country" plot on following website: https://covariants.org/variants/S.N501. This website pulls data from GISAID and makes it publicly accessible. This percentage will be accessed no sooner than 2021-04-12.
            \item Range:[0-100]
            \item Question URL:
            \url{https://pandemic.metaculus.com/questions/6477/-sn501-in-us-for-week-of-29-march/}
        \end{itemize}
\end{itemize}

\subsection{Survey 03 (Open on 2021-03-04, Close on 2021-03-15)}
\begin{itemize}
    \item Question 1
        \begin{itemize}
            \item Question: What will be the number of new incident US adult and pediatric admissions to the hospital with confirmed COVID-19 for the week beginning on 2021-03-21 and ending on 2021-03-27 (inclusive)?
            \item Resolution Criteria: This question will resolve as the total number of adult plus pediatric previous day admissions with confirmed COVID-19 as recorded in the Department of Health and Human Services report of COVID-19 reported patient impact and hospital capacity for the dates from 2021-03-22 to 2021-03-28, corresponding to the number of hospitalizations from 2021-03-21 to 2021-03-27. Daily updates are provided by the Department of Health and Human Services. The total previous day admissions is computed using two variables in this report: previous\_day\_admission\_adult\_covid\_confirmed and previous\_day\_admission\_pediatric\_covid\_confirmed and stored in Lehigh University's Computational Uncertainty Lab Github data repository. This report, and the resolution criteria, includes data on all 50 US states, Washington DC, Puerto Rico, and the US Virgin Islands (53 states and territories). The report will be accessed no sooner than (2021-04-04).
            \item Range:[0-80K]
            \item Metaculus question URL: \url{https://pandemic.metaculus.com/questions/6712/new-us-covid-hospital-admissions-21-27-march/}
            \item GJO question URL: \url{https://www.gjopen.com/questions/1952}
        \end{itemize}
    \item Question 2
        \begin{itemize}
            \item What will be the total number of new incident deaths due to COVID-19 in the US for the week beginning on 2021-03-21 and ending on 2021-03-27 (inclusive)?
            \item Resolution Criteria: This question will resolve as the number of new deaths due to confirmed COVID-19 for the week beginning on 2021-03-21 and ending on 2021-03-27 (inclusive) as recorded in the Johns Hopkins University (JHU) CSSE Github data repository. This file records the daily number of deaths by county. From this file deaths are summed across all counties and aggregated by week to generate the number of new deaths per week. The number of deaths for the week beginning on 2021-03-21 will be computed by adding the number of new deaths from the 2021-03-21 up to, and including, 2021-03-27. The report will be accessed no sooner than (2021-04-04).
            \item Range:[0-25k]
            \item Metaculus question URL: 
              \url{https://pandemic.metaculus.com/questions/6713/new-us-covid-deaths-21-27-march/}
            \item GJO question URL: \url{https://www.gjopen.com/questions/1953}
        \end{itemize}
    \item Question 3
        \begin{itemize}
            \item Question:What will be the number of new incident confirmed positive COVID-19 cases in the US beginning on 2021-03-21 and ending on 2021-03-27 (inclusive)?
            \item Resolution Criteria: This question will resolve as the number of new confirmed cases beginning on 2021-03-21 and ending on 2021-03-27 (inclusive) recorded in the Johns Hopkins University (JHU) CSSE Github data repository. This file records the daily number of cases by county. From this file cases are summed across all counties and aggregated by week to generate the number of new cases per week. The report will be accessed no sooner than 2021-04-04.
            \item Range:[0-2M]
            \item Metaculus question URL:
              \url{https://pandemic.metaculus.com/questions/6714/new-us-covid-cases-21-27-march/}
            \item GJO question URL: \url{https://www.gjopen.com/questions/1954}
        \end{itemize}
    \item Question 4
        \begin{itemize}
            \item Question: What will be the 7-day rolling average of the \% B.1.1.7 in the US on 27 March 2021 (between 21 March 2021 and 27 March 2021)?
            \item Resolution Criteria: This question will resolve as the 7 day rolling average of \%  sequences that are B.1.1.7 in the U.S. on 27 March 2021 (i.e. the average percentage between 21 March 2021 and 27 March 2021) at \href{https://outbreak.info/situation-reports?country=United\%20Kingdom&country=United\%20States&division=California&pango=B.1.1.7&selected=United\%20States&selectedType=country}{this website}. This percentage will be accessed no sooner than 6 April 2021.
            \item Range:[0-100]
            \item Metaculus question URL
            \url{https://pandemic.metaculus.com/questions/6717/-b117-in-the-us-21-27-march/}
        \end{itemize}
    \item Question 5 
        \begin{itemize}
            \item Question: In the context of community transmission, what will be the recommended minimum \% of positive COVID-19 cases that should be sequenced?
            \item Resolution Criteria: This question will resolve as the minimum CDC recommended percent of confirmed positive COVID-19 cases that should be sequenced that assumes community transmission. If the CDC does not release such guidance before the end of 2021, then the most-cited paper that provides a recommendation on the minimum recommended percent of positive COVID-19 cases that should be sequenced in the context of community transmission will be consulted on 1 January 2022.
            \item Range:[0-100]
            \item Metaculus question URL:
            \url{https://pandemic.metaculus.com/questions/6718/-covid-cases-that-should-be-sequenced/}
        \end{itemize}
    \item Question 6
        \begin{itemize}
            \item Question: How many variants of concern will be monitored by the US CDC as of 4 April?
            \item Resolution Criteria: This question will resolve as the number of variants of concern \href{https://www.cdc.gov/coronavirus/2019-ncov/transmission/variant-cases.html}{monitored by the CDC} as of Sunday, 2021-04-04. For example, as of 2021-03-02 this page shows that there are three variants: B.1.1.7, B.1.351, and P.1. This page is updated on Sundays, Tuesdays, and Thursdays by 7pm ET and will be accessed at approximately 10pm ET on 2021-04-04 (a Sunday).
            \item Range:[0-8]
            \item Metaculus question URL:
            \url{https://pandemic.metaculus.com/questions/6719/-variants-monitored-by-cdc-on-4-april/}
        \end{itemize}
    \item Question 7
        \begin{itemize}
            \item Question: What will be the cumulative number of people who receive one or more doses of a COVID-19 vaccine in the US on 2021-03-31?
            \item Resolution Criteria: This question will resolve as the cumulative number of people who receive one or more doses of a COVID-19 vaccine on 2021-03-31 as recorded by the Centers for Disease Control COVID-19 Data tracker. The radio buttons "People Receiving 1 or More Doses" and "Cumulative" will be selected and the bar corresponding to 2021-03-31 will be accessed. Data is updated daily by 8pm ET and will be accessed no sooner than 2021-04-04. If the CDC changes how it reports vaccination data, we will provide clarifying language as necessary. For purposes of this question, a person receiving a single-dose vaccine would count as a person having received one or more doses of a COVID-19 vaccine.
            \item Range:[0-140M]
            \item Metaculus question URL:
              \url{https://pandemic.metaculus.com/questions/6768/cumulative-1st-dose-us-vaccinations-31-march/}
            \item GJO question URL: \url{https://www.gjopen.com/questions/
                1955} 
        \end{itemize}
    \item Question 8
        \begin{itemize}
            \item Question: What will be the cumulative number of people who receive two doses of a COVID-19 vaccine in the US on 2021-03-31?
            \item Resolution Criteria: This question will resolve as the cumulative number of people who receive 2 doses of a COVID-19 vaccine on 2021-03-31 as recorded by the Centers for Disease Control COVID-19 Data tracker. The radio buttons "People Receiving 2 Doses" and "Cumulative" will be selected and the bar corresponding to 2021-03-31 will be accessed. Data is updated daily by 8pm ET and will be accessed no sooner than 2021-04-04. If the CDC changes how it reports vaccination data, we will provide clarifying language as necessary. For purposes of this question, a person receiving a single-dose vaccine would count as a person having received one or more doses of a COVID-19 vaccine. 
            \\ \\\textbf{Mar 8 edit:} On 2021-03-08, the CDC's vaccine tracker at \url{https://covid.cdc.gov/covid-data-tracker/#vaccinations} changed the "receiving 2 doses" figure to "fully vaccinated" to account for people who receive one dose of the Johnson \& Johnson vaccine, which has been authorized as a single-dose regimen (by contrast, Pfizer/BioNTech and Moderna are authorized as two-dose vaccines). This question will resolve on the basis of the new "fully vaccinated" figure reported by the CDC.
            \item Range:[0-70M]
            \item Metaculus question URL:
            \url{https://pandemic.metaculus.com/questions/6769/cumulative-two-dose-us-vaccinations-31-march/}
        \end{itemize}
\end{itemize}

\subsection{Survey 04 (Open on 2021-04-07, Close on 2021-04-20)}

\begin{itemize}
    \item Question 1
        \begin{itemize}
            \item Question: What will be the number of new incident US adult and pediatric admissions to the hospital with confirmed COVID-19 for the week beginning on 2021-04-25 and ending on 2021-05-01 (inclusive)?
            \item Resolution Criteria: This question will resolve as the total number of adult plus pediatric previous day admissions with confirmed COVID-19 as recorded in the Department of Health and Human Services report of COVID-19 reported patient impact and hospital capacity for the dates from 2021-04-25 to 2021-05-01, corresponding to the number of hospitalizations from 2021-04-25 to 2021-05-01. Daily updates are provided by the Department of Health and Human Services. The total previous day admissions is computed using two variables in this report: previous\_day\_admission\_adult\_covid\_confirmed and previous\_day\_admission\_pediatric\_covid\_confirmed and stored in Lehigh University's Computational Uncertainty Lab Github data repository. This report, and the resolution criteria, includes data on all 50 US states, Washington DC, Puerto Rico, and the US Virgin Islands (53 states and territories). The report will be accessed no sooner than (2021-09-05).
            \item Range:[0-75K]
            \item Metaculus URL: \url{https://pandemic.metaculus.com/questions/6985/new-us-covid-hospital-admissions-25-apr-1-may/}
            \item GJO URL: \url{https://www.gjopen.com/questions/1976}
        \end{itemize}
    \item Question 2
        \begin{itemize}
            \item What will be the total number of new incident deaths due to COVID-19 in the US for the week beginning on 2021-04-25 and ending on 2021-05-01 (inclusive)?
            \item Resolution Criteria: This question will resolve as the number of new deaths due to confirmed COVID-19 for the week beginning on 2021-04-25 and ending on 2021-05-01 (inclusive) as recorded in the Johns Hopkins University (JHU) CSSE Github data repository. This file records the daily number of deaths by county. From this file deaths are summed across all counties and aggregated by week to generate the number of new deaths per week. The number of deaths for the week beginning on 2021-04-25 will be computed by adding the number of new deaths from the 2021-04-25 up to, and including, 2021-05-01. The report will be accessed no sooner than 9 May 2021.
            \item Range:[0-10k]
            \item Metaculus URL: 
            \url{https://pandemic.metaculus.com/questions/6986/new-us-covid-deaths-25-apr-1-may/}
            \item GJO URL: \url{https://www.gjopen.com/questions/1977}
        \end{itemize}
    \item Question 3
        \begin{itemize}
            \item Question: What will be the number of new incident confirmed positive COVID-19 cases in the US beginning on 2021-04-25 and ending on 2021-05-01 (inclusive)?
            \item Resolution Criteria: This question will resolve as the number of new confirmed cases beginning on 2021-04-25 and ending on 2021-05-01 (inclusive) recorded in the Johns Hopkins University (JHU) CSSE Github data repository. This file records the daily number of cases by county. From this file cases are summed across all counties and aggregated by week to generate the number of new cases per week. The report will be accessed no sooner than 2021-05-09.
            \item Range:[0-1.2M]
            \item Metaculus URL:
            \url{https://pandemic.metaculus.com/questions/6987/new-us-covid-cases-25-apr-1-may/}
            \item GJO URL: \url{https://www.gjopen.com/questions/1978}
        \end{itemize}
    \item Question 4
        \begin{itemize}
            \item Question: What will be the cumulative number of people who receive one or more doses of a COVID-19 vaccine in the US on 2021-04-30?
            \item Resolution Criteria: This question will resolve as the cumulative number of people who have received one or more doses of a vaccine on 2021-04-30 as recorded by the \href{https://covid.cdc.gov/covid-data-tracker/#vaccination-trends}{Centers for Disease Control COVID-19 Data tracker under Vaccine Trends}. The dashboard is updated daily at 8pm ET and will be accessed no sooner than 9 May 2021.
            \item Range:[109M-215M]
            \item Metaculus URL
            \url{https://pandemic.metaculus.com/questions/6988/cumulative-1st-dose-us-vaccinations-30-april/}
            \item GJO URL: \url{https://www.gjopen.com/questions/1979}
        \end{itemize}
    \item Question 5 
        \begin{itemize}
            \item Question: What will be the cumulative number of people who are fully vaccinated against COVID-19 in the US on 2021-04-30?
            \item Resolution Criteria: This question will resolve as the cumulative number of people who receive one or more doses of a COVID-19 vaccine on 2021-04-30 as recorded by the \href{https://covid.cdc.gov/covid-data-tracker/#vaccination-trends}{Centers for Disease Control COVID-19 Data tracker} in the column "People Fully Vaccinated". The dashboard is updated daily at 8pm ET and will be accessed no sooner than 9 May 2021.
            \item Range:[64M-130M]
            \item Metaculus URL:
            \url{https://pandemic.metaculus.com/questions/6989/cumulative-fully-vaccinated-in-us-on-30-april/}
            \item GJO URL: \url{https://www.gjopen.com/questions/1980}
        \end{itemize}
    \item Question 6
        \begin{itemize}
            \item Question: What will be the cumulative number of deaths in the US due to COVID-19 on 2021-12-31?
            \item Resolution Criteria: This question will resolve as the number of cumulative deaths due to confirmed COVID-19 on 2021-12-31 as recorded in the \href{https://github.com/CSSEGISandData/COVID-19/blob/master/csse_covid_19_data/csse_covid_19_time_series/time_series_covid19_deaths_US.csv}{Johns Hopkins University (JHU) CSSE Github data repository}. This file records the daily number of deaths by county. The number of cumulative deaths at the end of the year will be computed by adding the cumulative number of deaths across states. This data, and the resolution criteria, includes data on all 50 US states, Washington DC, Puerto Rico, and the US Virgin Islands (53 states and territories). The report will be accessed no sooner than 9 January 2022.
            \item Range:[555K-1.5M]
            \item Metaculus URL:
            \url{https://pandemic.metaculus.com/questions/6990/cumulative-us-covid-deaths-by-end-of-2021/}
        \end{itemize}
    \item Question 7
        \begin{itemize}
            \item Question: What will be the 7-day rolling average of \% B.1.1.7 in the US on 30 Apr 2021 (between 24 Apr 2021 and 30 Apr 2021)?
            \item Resolution Criteria: This question will resolve as the 7-day rolling average of \% B.1.1.7 in the US (in other words, the frequency of B.1.1.7 as a percentage of all sequenced SARS-CoV-2 cases) on 30 Apr 2021 (the average percentage between 24 Apr 2021 and 30 Apr 2021) at the following website: \url{https://outbreak.info/situation-reports?pango=B.1.1.7&loc=USA&selected=USA}. This percentage will be accessed no sooner than 9 May 2021.
            \item Range:[0-100]
            \item Metaculus URL:
            \url{https://pandemic.metaculus.com/questions/6991/-b117-in-us-on-30-april/}
        \end{itemize}
    \item Question 8
        \begin{itemize}
            \item Question:On what date will the United States CDC announce that they are tracking a SARS-CoV-2 variant that they classify as a variant of high consequence (VOHC)?
            \item Resolution Criteria: This question will resolve as the date that a SARS-CoV-2 variant is categorized under the "Variant of High Consequence" section on the CDC's \href{https://www.cdc.gov/coronavirus/2019-ncov/cases-updates/variant-surveillance/variant-info.html}{SARS-CoV-2 Variant Classifications and Definitions page}.

            If no variant is classified as a VOHC before 1 January 2023, then this resolves as > 31 December 2022.
            \item Range:[2021-04-07 - $>$2022-12-31]
            \item Metaculus URL:
            \url{https://pandemic.metaculus.com/questions/6992/variant-of-high-consequence-before-2023}
        \end{itemize}
\end{itemize}

\subsection{Survey 05 (Open on 2021-05-05, Close on 2021-05-18)}

\begin{itemize}
    \item Question 1
        \begin{itemize}
            \item Question: What will be the number of new incident U.S. adult and pediatric admissions to the hospital with confirmed COVID-19 for the week beginning on 2021-05-30 and ending on 2021-06-05 (inclusive)?
            \item Range:$[0-60K]$
            \item Metaculus URL:
            \url{https://pandemic.metaculus.com/questions/7156/new-us-hospital-admissions-30-may-5-june/}
            \item GJO URL: \url{https://www.gjopen.com/questions/1996}
        \end{itemize}
    \item Question 2
        \begin{itemize}
            \item What will be the total number of new incident deaths due to COVID-19 in the US for the week beginning on 2021-05-30 and ending on 2021-06-05 (inclusive)?
            \item Resolution Criteria: This question will resolve as the number of new deaths due to confirmed COVID-19 for the week beginning on 2021-05-30 and ending on 2021-06-05 (inclusive) as recorded in the Johns Hopkins University (JHU) CSSE Github data repository. This file records the daily number of deaths by county. From this file deaths are summed across all counties and aggregated by week to generate the number of new deaths per week. The number of deaths for the week beginning on 2021-05-30 will be computed by adding the number of new deaths from the 2021-05-30 up to, and including, 2021-06-05. The report will be accessed no sooner than 2021-06-14.
            \item Range:$[0-8K]$
            \item Metaculus URL: 
            \url{https://pandemic.metaculus.com/questions/7157/new-us-covid-deaths-30-may-5-june/}
            \item GJO URL: \url{https://www.gjopen.com/questions/1997}
        \end{itemize}
    \item Question 3
        \begin{itemize}
            \item Question: What will be the number of new incident confirmed positive COVID-19 cases in the US beginning on 2021-05-30 and ending on 2021-06-05 (inclusive)?
            \item Resolution Criteria: This question will resolve as the number of new confirmed cases beginning on 2021-05-30 and ending on 2021-06-05 (inclusive) recorded in the Johns Hopkins University (JHU) CSSE Github data repository. This file records the daily number of cases by county. From this file cases are summed across all counties and aggregated by week to generate the number of new cases per week. The report will be accessed on 2021-06-14.
            \item Range:$[0-800K]$
            \item Metaculus URL:
            \url{https://pandemic.metaculus.com/questions/7158/new-us-covid-cases-30-may-5-june/}
            \item GJO URL: \url{https://www.gjopen.com/questions/1998}
        \end{itemize}
    \item Question 4
        \begin{itemize}
            \item Question: What will be the cumulative number of people who receive one or more doses of a COVID-19 vaccine in the US on 2021-05-31?
            \item Resolution Criteria: This question will resolve as the cumulative number of people who have received one or more doses of a vaccine on 2021-05-31 as recorded by the \href{https://covid.cdc.gov/covid-data-tracker/#vaccination-trends}{Centers for Disease Control COVID-19 Data tracker under Vaccine Trends}. The dashboard is updated daily at 8pm ET and will be accessed no sooner than 2021-06-14
            \item Range:$[148M-185M]$
            \item Metaculus URL
            \url{https://pandemic.metaculus.com/questions/7159/cumulative-1st-dose-us-vaccinations-31-may/}
            \item GJO URL: \url{https://www.gjopen.com/questions/1999}
        \end{itemize}
    \item Question 5 
        \begin{itemize}
            \item Question: What will be the cumulative number of people who are fully vaccinated against COVID-19 in the US on 2021-05-31?
            \item Resolution Criteria: This question will resolve as the cumulative number of people who receive one or more doses of a COVID-19 vaccine on 2021-05-31 as recorded by the \href{https://covid.cdc.gov/covid-data-tracker/#vaccination-trends}{Centers for Disease Control COVID-19 Data tracker} in the column "People Fully Vaccinated". The dashboard is updated daily at 8pm ET and will be accessed no sooner than 2021-06-14.
            \item Range:$[106M-165M]$
            \item Metaculus URL:
            \url{https://pandemic.metaculus.com/questions/7160/cumulative-fully-vaccinated-in-us-on-31-may/}
            \item GJO URL: \url{https://www.gjopen.com/questions/2000}
        \end{itemize}
    \item Question 6
        \begin{itemize}
            \item Question: What will be the cumulative number of deaths in the US due to COVID-19 on 2021-12-31?
            \item Resolution Criteria: This question will resolve as the number of cumulative deaths due to confirmed COVID-19 on 2021-12-31 as recorded in the \href{https://github.com/CSSEGISandData/COVID-19/blob/master/csse_covid_19_data/csse_covid_19_time_series/time_series_covid19_deaths_US.csv}{Johns Hopkins University (JHU) CSSE Github data repository}. This file records the daily number of deaths by county. The number of cumulative deaths at the end of the year will be computed by adding the cumulative number of deaths across states. This data, and the resolution criteria, includes data on all 50 US states, Washington DC, Puerto Rico, and the US Virgin Islands (53 states and territories). The report will be accessed no sooner than 2022-01-09.
            \item Range:$[575K-1.1M]$
            \item Metaculus URL:
            \url{https://pandemic.metaculus.com/questions/7161/cumulative-us-covid-deaths-by-end-of-2021/}
        \end{itemize}
    \item Question 7
        \begin{itemize}
            \item Question: What will be the \% prevalence of SARS-CoV-2 variants thought to partially escape immunity for the two-week period 23 May - 05 Jun 2021?
            \item Resolution Criteria: This question will resolve on the basis of the first update that shows figures for the two-week period ending 05 Jun of the "Weighted Estimates of Proportions of SARS-CoV-2 Lineages" table on the U.S. CDC's \href{https://covid.cdc.gov/covid-data-tracker/#variant-proportions}{"Variant Proportions"} page. The percentages of variants that cause "reduced neutralization by convalescent and post-vaccination sera" will be added up. If between now and 05 Jun there are additional variants classified by the CDC as variants that cause "reduced neutralization by convalescent and post-vaccination sera," these will count toward the total percent figure. Likewise, if any of the variants that are currently classified as causing partial immune escape are removed from being classified as such, they will no longer count toward the total percent figure.
            \item Range:$[0-100]$
            \item Metaculus URL:
            \url{https://pandemic.metaculus.com/questions/7164/prevalence-of-immune-evading-variants-5-june/}
        \end{itemize}
    \item Question 8
        \begin{itemize}
            \item Question:On what date will the United States CDC announce that they are tracking a SARS-CoV-2 variant that they classify as a variant of high consequence (VOHC)?
            \item Resolution Criteria: This question will resolve as the date that a SARS-CoV-2 variant is categorized under the "Variant of High Consequence" section on the CDC's \href{https://www.cdc.gov/coronavirus/2019-ncov/cases-updates/variant-surveillance/variant-info.html}{SARS-CoV-2 Variant Classifications and Definitions page}. If no variant is classified as a VOHC before 1 January 2025, then this resolves as $>$ 31 December 2024.\item Range:$[2021-05-01 - >2024-12-31]$
            \item Metaculus URL:
            \url{https://pandemic.metaculus.com/questions/7163/variant-of-high-consequence-before-2025/}
        \end{itemize}
\end{itemize}

\subsection{Survey 6 (Open on 2021-06-02, Close on 2021-06-16)}
\begin{itemize}
    \item Question 1
        \begin{itemize}
            \item Question: What will be the number of new incident   U.S. adult and pediatric admissions to the hospital with confirmed COVID-19   for the week beginning on 2021-06-27 and ending on 2021-07-03 (inclusive)? 
            \item Resolution Criteria: This question will resolve as the total   number of adult plus pediatric previous day admissions with confirmed   COVID-19 as recorded in the Department of Health and Human Service report of   COVID-19 reported patient impact and hospital capacity for the dates from   2021-06-28 to 2021-07-04, corresponding to the number of hospitalizations   from 2021-06-27 to 2021-07-03. Daily updates are provided by the Department   of Health and Human Services. The total previous day admissions is computed   using two variables in this report: previous\_day\_admission\_adult\_covid\_confirmed   and previous\_day\_admission\_pediatric\_covid\_confirmed and stored in Lehigh   University's Computational Uncertainty Lab Github data repository. This   report, and the resolution criteria, includes data on all 50 US states,   Washington DC, Puerto Rico, and the US Virgin Islands (53 states and   territories). The report will be accessed on 2021-07-12. 
            \item Range:[0-40K]
            \item Metaculus URL: \url{https://pandemic.metaculus.com/questions/7301/new-us-hospital-admissions-27-june-3-july/}
            \item GJO URL: \url{https://www.gjopen.com/questions/2023}
        \end{itemize}
    \item Question 2
        \begin{itemize}
            \item What will be the total number of new   incident deaths due to COVID-19 in the US for the week beginning on 2021-06-27 and ending on 2021-07-03 (inclusive)?  
            \item Resolution Criteria: This question will resolve as the number of new deaths due to confirmed COVID-19 for the week beginning on 2021-06-27 and ending on 2021-07-03 (inclusive) as recorded in the Johns Hopkins University (JHU) CSSE Github data repository. This file records the daily number of deaths by county. From this file deaths are summed across all counties and aggregated by week to generate the number of new deaths per week. The number of deaths for the week beginning on 2021-06-27 will be computed by adding the number of new deaths from the 2021-06-27 up to, and including, 2021-07-03. The report will be accessed on 2021-07-12.
            \item Range:[0-5k]
            \item Metaculus URL: 
            \url{https://pandemic.metaculus.com/questions/7302/new-us-covid-deaths-27-june-3-july/}
            \item GJO URL: \url{https://www.gjopen.com/questions/2024}
        \end{itemize}
    \item Question 3
        \begin{itemize}
            \item Question: What will be the number of new incident confirmed positive COVID-19 cases in the US beginning on 2021-06-27 and ending on 2021-07-03 (inclusive)?
            \item Resolution Criteria: This question will resolve as the number of new confirmed cases beginning on 2021-06-27 and ending on 2021-07-03 (inclusive) recorded in the Johns Hopkins University (JHU) CSSE Github data repository. This file records the daily number of cases by county. From this file cases are summed across all counties and aggregated by week to generate the number of new cases per week. The report will be accessed no sooner than 2021-07-12.
            \item Range:[0-300K]
            \item Metaculus URL:
            \url{https://pandemic.metaculus.com/questions/7303/new-us-covid-cases-27-june-3-july/}
            \item GJO URL: \url{https://www.gjopen.com/questions/2025}
        \end{itemize}
    \item Question 4
        \begin{itemize}
            \item Question: What will be the cumulative number of people who receive one or more doses of a COVID-19 vaccine in the US on 2021-06-30?
            \item Resolution Criteria: This question will resolve as the   cumulative number of people who have received one or more doses of a vaccine   on 2021-06-30 as recorded by the Centers for Disease Control COVID-19 Data   tracker under Vaccine Trends. The dashboard is updated daily at 8pm ET and will   be accessed no sooner than 2021-07-12.
            \item Range:[169M-200M]
            \item Metaculus URL
            \url{https://pandemic.metaculus.com/questions/7305/cumulative-1st-dose-us-vaccinations-30-june/}
            \item GJO URL: \url{https://www.gjopen.com/questions/2026}
        \end{itemize}
    \item Question 5 
        \begin{itemize}
            \item Question: What will be the cumulative number of people who are fully vaccinated against COVID-19 in the US on 2021-06-30?
            \item Resolution Criteria: This question will resolve as the   cumulative number of people who receive one or more doses of a COVID-19   vaccine on 2021-06-30 as recorded by the Centers for Disease Control COVID-19   Data tracker in the column People Fully Vaccinated. The dashboard is   updated daily at 8pm ET and will be accessed no sooner than 2021-07-12.
            \item Range:[136M-180M]
            \item Metaculus URL:
            \url{https://pandemic.metaculus.com/questions/7306/cumulative-fully-vaccinated-in-us-on-30-june/ }
            \item GJO URL: \url{https://www.gjopen.com/questions/2027}
        \end{itemize}
    \item Question 6
        \begin{itemize}
            \item Question: What will be the cumulative number of deaths in the US due to COVID-19 on 2021-12-31?
            \item Resolution Criteria: This question will resolve as the number of cumulative deaths due to confirmed COVID-19 on 2021-12-31 as recorded in the \href{https://github.com/CSSEGISandData/COVID-19/blob/master/csse_covid_19_data/csse_covid_19_time_series/time_series_covid19_deaths_US.csv}{Johns Hopkins University (JHU) CSSE Github data repository}. This file records the daily number of deaths by county. The number of cumulative deaths at the end of the year will be computed by adding the cumulative number of deaths across states. This data, and the resolution criteria, includes data on all 50 US states, Washington DC, Puerto Rico, and the US Virgin Islands (53 states and territories). The report will be accessed no sooner than 9 January 2022.
            \item Range:[595K-1.2M]
            \item Metaculus URL:
            \url{https://pandemic.metaculus.com/questions/7307/cumulative-us-covid-deaths-by-end-of-2021/}
        \end{itemize}
    \item Question 7
        \begin{itemize}
            \item Question: What will be the prevalence of SARS-CoV-2 variants thought to partially escape immunity for the two-week period 20 June - 03 July 2021?
            \item Resolution Criteria: This question will resolve on the basis of the first update that shows figures for the two-week period ending 03 July   of the "Weighted Estimates of Proportions of SARS-CoV-2 Lineages" table on the U.S. CDC's "Variant Proportions" page. The percentages of variants that cause "reduced neutralization by convalescent and post-vaccination sera" will be added up. If between now and 03 July there are additional variants classified by the CDC as variants that cause "reduced neutralization" by convalescent and/or post-vaccination sera, these will count toward the total percent figure. Likewise, if any of the variants that are currently classified as causing partial immune escape are removed from being classified as such, they will no longer count toward the total percent figure. 
            \item Range:[0-100]
            \item Metaculus URL:
            \url{https://pandemic.metaculus.com/questions/7308/prevalence-of-immune-evading-variants-3-july/}
        \end{itemize}
\end{itemize}

\clearpage
\section{Forecasting platforms}
\label{supp.forecastingplat}

\setcounter{figure}{0}

\begin{figure}[ht!]
    \centering
    \fbox{\includegraphics[scale=0.625,angle=90]{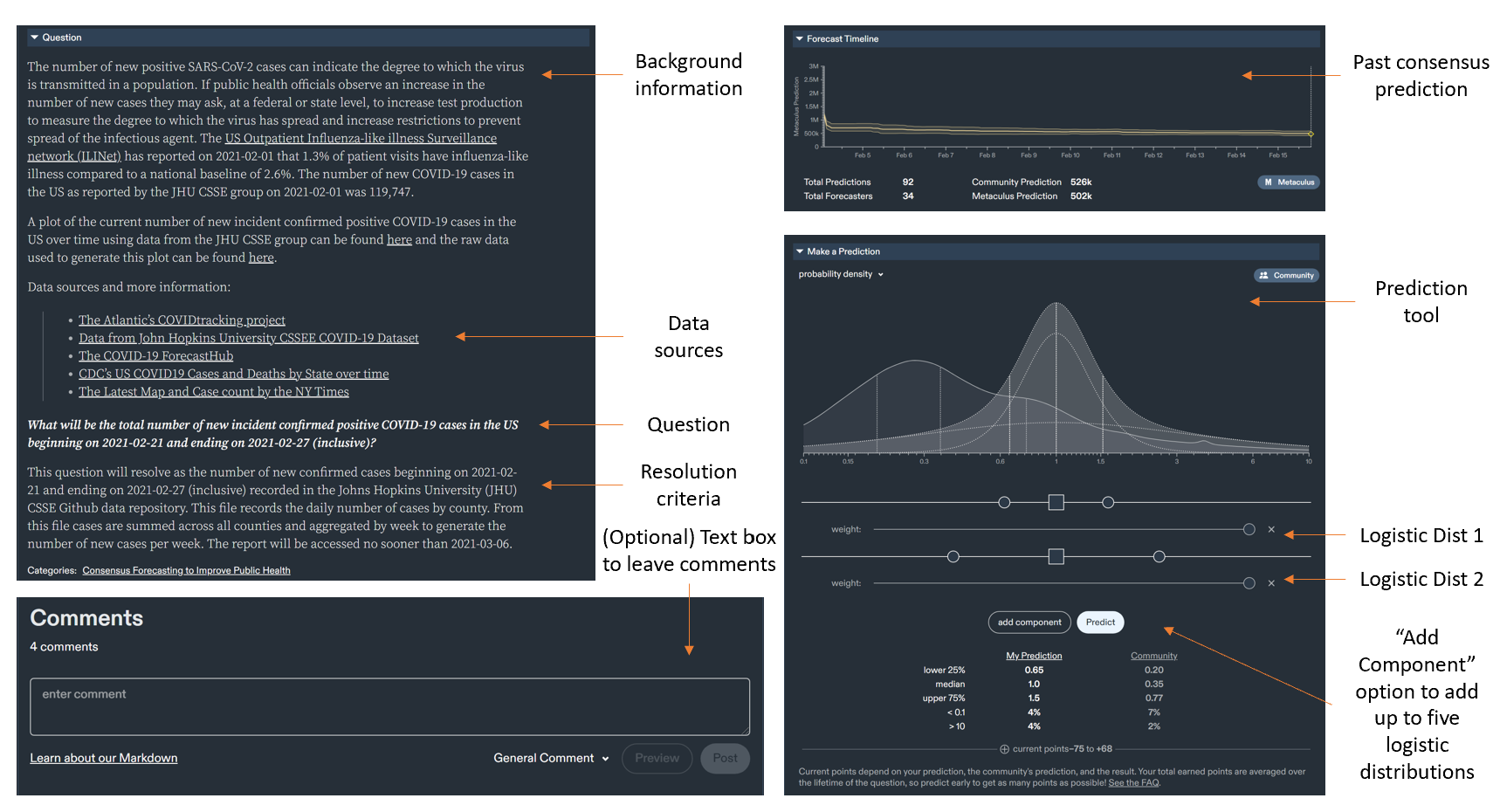}}
    \caption{The Metaculus format for submitting a forecast over a target of interest. Forecasters are presented with background information about the target and data sources that may be helpful when building a forecast. A forecaster can also view a consensus forecast from when the question was posed until present. A clearly defined question is asked in bold font and forecasters are also presented with the resolution criteria, how the ground truth for this question will be determined.
    Forecasters build a predictive distribution as a mixture of five logistic distributions. Forecasters, if they wish, can also share useful comments and questions with others in a chat box underneath their forecast.~\label{fig.metac}}
\end{figure}

\begin{figure}[ht!]
    \centering
    \fbox{\includegraphics{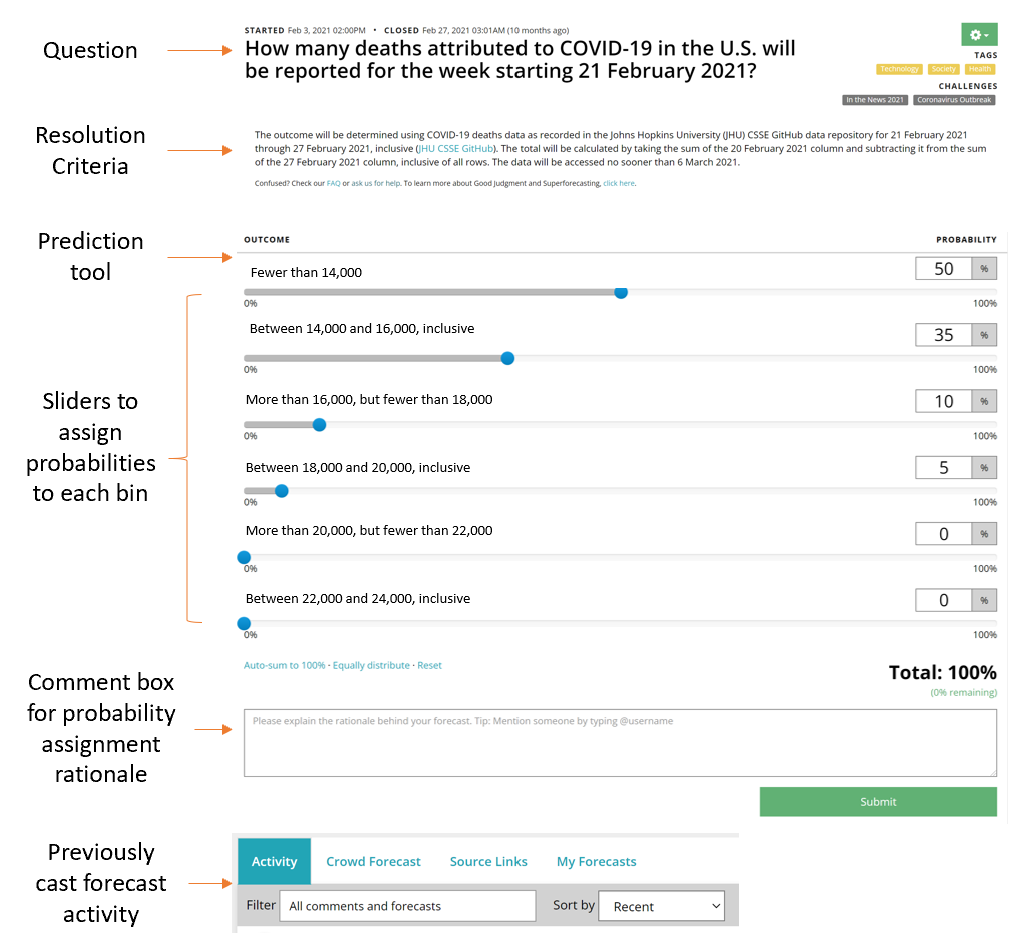}}
    \caption{The Good Judgment Open format for submitted a forecast over a target of interest. The question is posed to forecasters at the top of the page in bold font and underneath the question is the resolution criteria describing how the ground truth for this question will be determined. Forecasters can also view the current and past consensus distribution for this question. Forecasters build their predictive distribution by sliding $N$ bars which represent intervals over possible truth values. Forecasters, if they wish, can also share useful comments and questions with others~\label{fig.gjo}}
\end{figure}

\clearpage
\section{List of included computational models form the COVID-19 Forecast Hub}
\label{tab.listOfCompModels}


\begin{table}[ht!]
  \begin{tabular}{lp{13cm}}
Model & Link to Data \\
    \hline
    BPagano-RtDriven &  \url{https://github.com/reichlab/covid19-forecast-hub/tree/master/data-processed/BPagano-RtDriven} \\
    CEID-Walk & \url{https://github.com/reichlab/covid19-forecast-hub/tree/master/data-processed/CEID-Walk} \\
    Columbia\_UNC-SurvCon & \url{https://github.com/reichlab/covid19-forecast-hub/tree/master/data-processed/Columbia_UNC-SurvCon} \\
    Covid19Sim-Simulator & \url{https://github.com/reichlab/covid19-forecast-hub/tree/master/data-processed/Covid19Sim-Simulator} \\
    CovidAnalytics-DELPHI & \url{https://github.com/reichlab/covid19-forecast-hub/tree/master/data-processed/CovidAnalytics-DELPHI} \\
    COVIDhub-baseline & \url{https://github.com/reichlab/covid19-forecast-hub/tree/master/data-processed/COVIDhub-baseline} \\
    CU-select & \url{https://github.com/reichlab/covid19-forecast-hub/tree/master/data-processed/CU-select} \\
    DDS-NBDS & \url{https://github.com/reichlab/covid19-forecast-hub/tree/master/data-processed/DDS-NBDS} \\
    IEM\_MED-CovidProject & \url{https://github.com/reichlab/covid19-forecast-hub/tree/master/data-processed/IEM_MED-CovidProject} \\
    IowaStateLW-STEM & \url{https://github.com/reichlab/covid19-forecast-hub/tree/master/data-processed/IowaStateLW-STEM} \\
    JHU\_IDD-CovidSP & \url{https://github.com/reichlab/covid19-forecast-hub/tree/master/data-processed/JHU_IDD-CovidSP} \\
    JHUAPL-Bucky  & \url{https://github.com/reichlab/covid19-forecast-hub/tree/master/data-processed/JHUAPL-Bucky} \\
    Karlen-pypm & \url{https://github.com/reichlab/covid19-forecast-hub/tree/master/data-processed/Karlen-pypm} \\
    LANL-GrowthRate & \url{https://github.com/reichlab/covid19-forecast-hub/tree/master/data-processed/LANL-GrowthRate} \\
    LNQ-ens1 & \url{https://github.com/reichlab/covid19-forecast-hub/tree/master/data-processed/LNQ-ens1} \\
    Microsoft-DeepSTIA & \url{https://github.com/reichlab/covid19-forecast-hub/tree/master/data-processed/Microsoft-DeepSTIA} \\
    OneQuietNight-ML & \url{https://github.com/reichlab/covid19-forecast-hub/tree/master/data-processed/OneQuietNight-ML} \\
    QJHong-Encounter & \url{https://github.com/reichlab/covid19-forecast-hub/tree/master/data-processed/QJHong-Encounter} \\
    RobertWalraven-ESG & \url{https://github.com/reichlab/covid19-forecast-hub/tree/master/data-processed/RobertWalraven-ESG} \\
    SigSci-TS & \url{https://github.com/reichlab/covid19-forecast-hub/tree/master/data-processed/SigSci-TS} \\

  \end{tabular}
  \end{table}

    \begin{table}[ht!]
      \begin{tabular}{lp{13cm}}
           UCF-AEM & \url{https://github.com/reichlab/covid19-forecast-hub/tree/master/data-processed/UCF-AEM} \\
    UCLA-SuEIR & \url{https://github.com/reichlab/covid19-forecast-hub/tree/master/data-processed/UCLA-SuEIR} \\
    UMich-RidgeTfReg & \url{https://github.com/reichlab/covid19-forecast-hub/tree/master/data-processed/UMich-RidgeTfReg} \\
    UpstateSU-GRU & \url{https://github.com/reichlab/covid19-forecast-hub/tree/master/data-processed/UpstateSU-GRU} \\
    USC-SI\_kJalpha & \url{https://github.com/reichlab/covid19-forecast-hub/tree/master/data-processed/USC-SI_kJalpha} \\
    UVA-Ensemble & \url{https://github.com/reichlab/covid19-forecast-hub/tree/master/data-processed/UVA-Ensemble} \\
    JHU\_CSSE-DECOM & \url{https://github.com/reichlab/covid19-forecast-hub/tree/master/data-processed/JHU_CSSE-DECOM} \\
    MIT\_ISOLAT-Mixtures & \url{https://github.com/reichlab/covid19-forecast-hub/tree/master/data-processed/MIT_ISOLAT-Mixtures} \\
    MOBS-GLEAM\_COVID & \url{https://github.com/reichlab/covid19-forecast-hub/tree/master/data-processed/MOBS-GLEAM_COVID} \\
    FRBSF\_Wilson-Econometric & \url{https://github.com/reichlab/covid19-forecast-hub/tree/master/data-processed/FRBSF_Wilson-Econometric} \\
    MUNI-ARIMA & \url{https://github.com/reichlab/covid19-forecast-hub/tree/master/data-processed/MUNI-ARIMA} \\
    MIT-Cassandra & \url{https://github.com/reichlab/covid19-forecast-hub/tree/master/data-processed/MIT-Cassandra} \\
    epiforecasts-ensemble1 & \url{https://github.com/reichlab/covid19-forecast-hub/tree/master/data-processed/epiforecasts-ensemble1} \\
    GT-DeepCOVID & \url{https://github.com/reichlab/covid19-forecast-hub/tree/master/data-processed/GT-DeepCOVID} \\
    MIT\_CritData-GBCF & \url{https://github.com/reichlab/covid19-forecast-hub/tree/master/data-processed/MIT_CritData-GBCF} \\
    MITCovAlliance-SIR & \url{https://github.com/reichlab/covid19-forecast-hub/tree/master/data-processed/MITCovAlliance-SIR} \\
    OliverWyman-Navigator & \url{https://github.com/reichlab/covid19-forecast-hub/tree/master/data-processed/OliverWyman-Navigator} \\
    PSI-DRAFT & \url{https://github.com/reichlab/covid19-forecast-hub/tree/master/data-processed/PSI-DRAFT} \\
    RPI\_UW-Mob\_Collision & \url{https://github.com/reichlab/covid19-forecast-hub/tree/master/data-processed/RPI_UW-Mob_Collision} \\
    SteveMcConnell-CovidComplete & \url{https://github.com/reichlab/covid19-forecast-hub/tree/master/data-processed/SteveMcConnell-CovidComplete} \\
    UA-EpiCovDA &  \url{https://github.com/reichlab/covid19-forecast-hub/tree/master/data-processed/UA-EpiCovDA} \\
    UCM\_MESALab-FoGSEIR & \url{https://github.com/reichlab/covid19-forecast-hub/tree/master/data-processed/UCM_MESALab-FoGSEIR} \\
    UCSD\_NEU-DeepGLEAM & \url{https://github.com/reichlab/covid19-forecast-hub/tree/master/data-processed/UCSD_NEU-DeepGLEAM} \\
    UMass-MechBayes & \url{https://github.com/reichlab/covid19-forecast-hub/tree/master/data-processed/UMass-MechBayes} \\
    UT-Mobility & \url{https://github.com/reichlab/covid19-forecast-hub/tree/master/data-processed/UT-Mobility} \\
    IHME-CurveFit & \url{https://github.com/reichlab/covid19-forecast-hub/tree/master/data-processed/IHME-CurveFit} \\
    \hline

  \end{tabular}
  \end{table}

\clearpage
\section{Paired difference in WIS between a performance based and equally weighted ensemble across surveys}

\begin{figure}[ht!]
    \centering
    \includegraphics[scale=0.90]{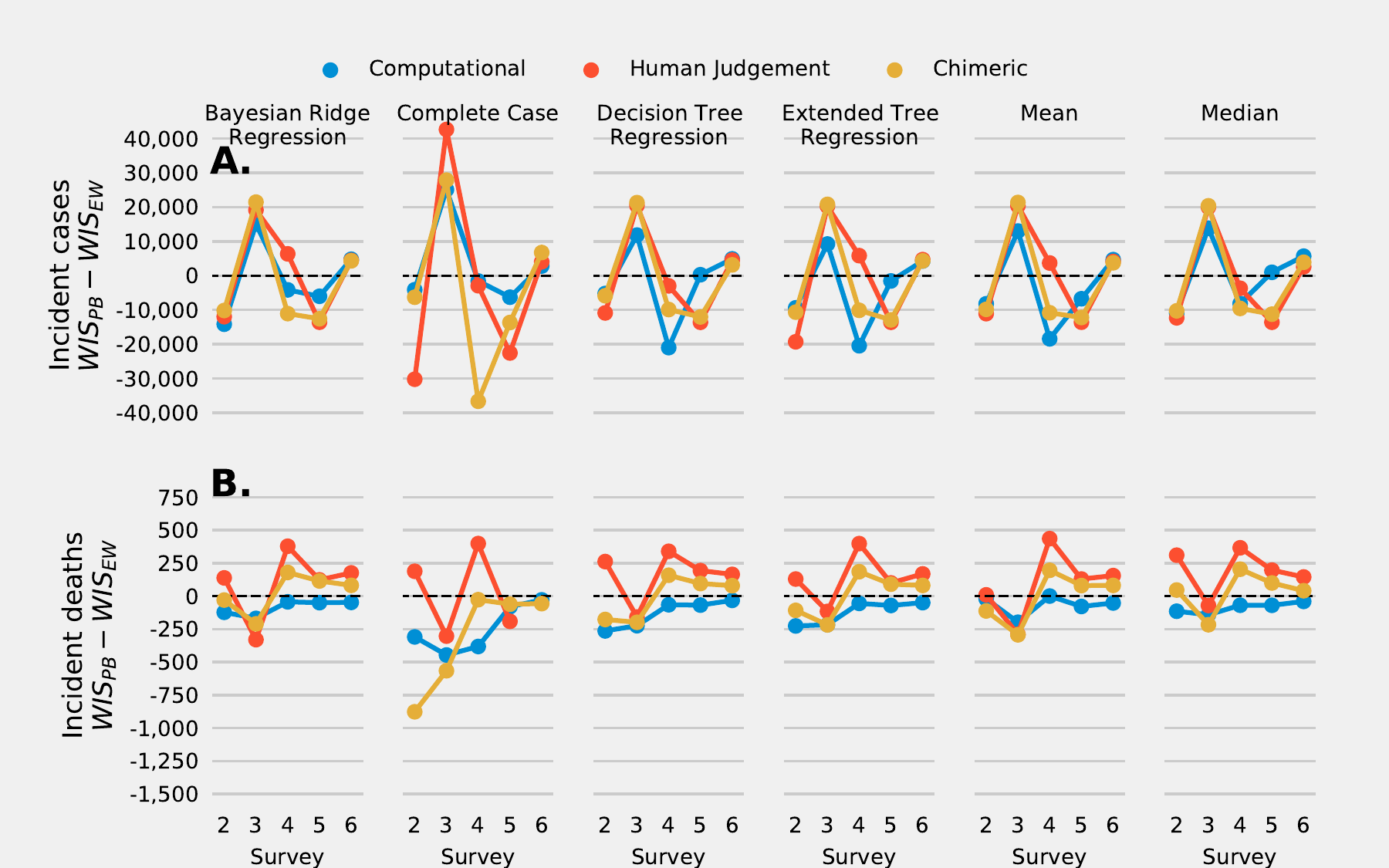}
    \caption{Paired difference in WIS score for predictions of incident cases~(A.) and deaths~(B.) for surveys 2, 3, 4, 5, 6 between a performance based and equally weighted computational ensemble~(blue), human judgement~(red), and chimeric ensemble~(gold) using a "spotty memory" imputation approach. 
    ~\label{fig.pairedOverSurveys_spotty}}
\end{figure}

\begin{figure}[ht!]
    \centering
    \includegraphics[scale=0.90]{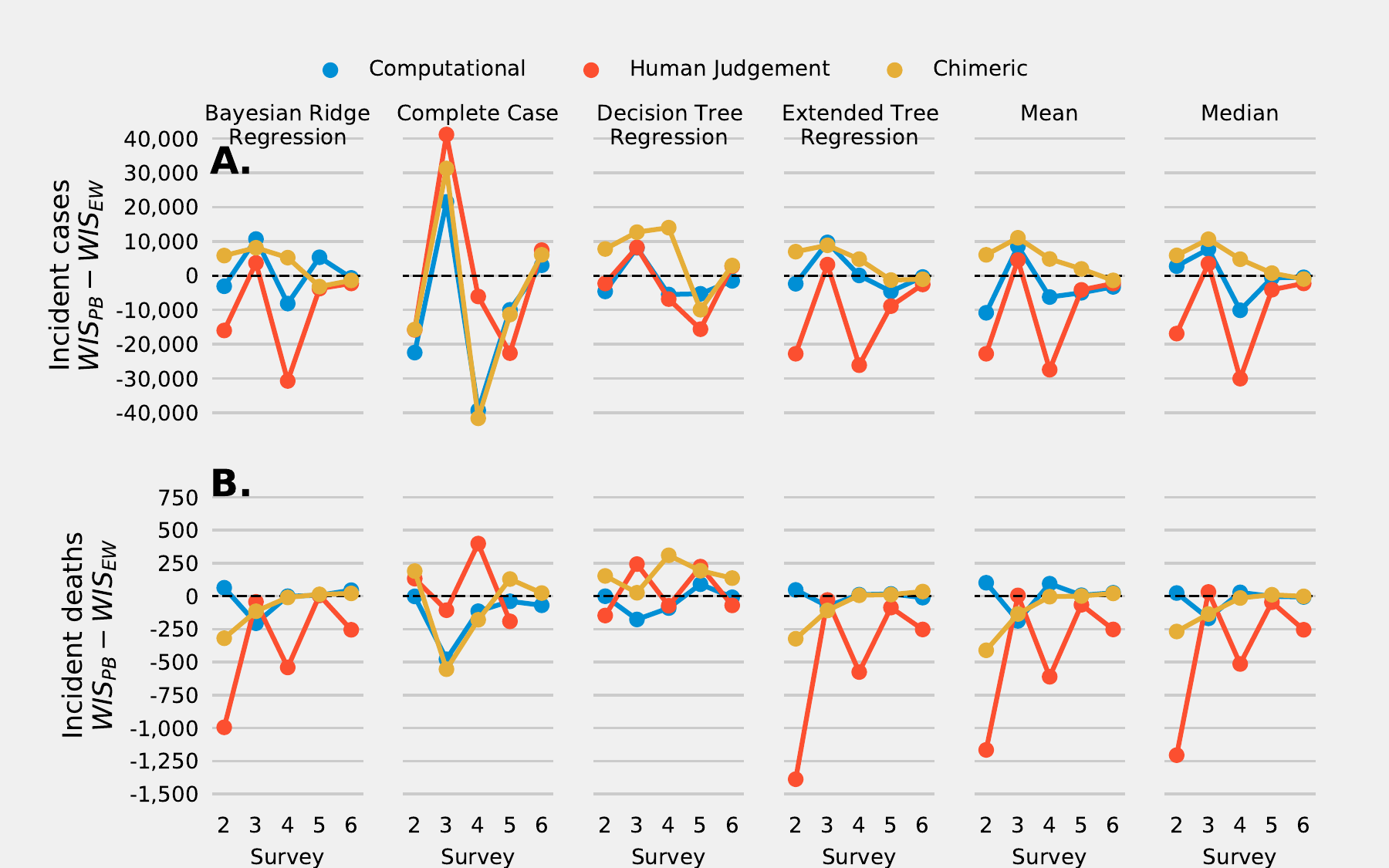}
    \caption{Paired difference in WIS score for predictions of incident cases~(A.) and deaths~(B.) for surveys 2, 3, 4, 5, 6 between a performance based and equally weighted computational ensemble~(blue), human judgement~(red), and chimeric ensemble~(gold) using a "defer to the crowd" imputation approach. 
    ~\label{fig.pairedOverSurveys_defer}}
\end{figure}

\clearpage
\section{Counts of computational and human judgement predictions}

\begin{table}[ht!]
    \centering
    \begin{tabular}{ccccc}
       \hline
       Survey & Target & Computational models & Metaculus & GJO  \\
       \hline
       1 & Cases & 25 & 22 & 00 \\
       2 & Cases & 27 & 14 & 35 \\
       3 & Cases & 27 & 20 & 40 \\
       4 & Cases & 24 & 06 & 32 \\
       5 & Cases & 24 & 15 & 21 \\
       6 & Cases & 23 & 14 & 19 \\
       
       1 & Deaths & 34 & 24  & 00 \\
       2 & Deaths & 34 & 13  & 42 \\
       3 & Deaths & 34 & 22  & 39 \\
       4 & Deaths & 34 & 05  & 34 \\
       5 & Deaths & 33 & 16  & 24 \\
       6 & Deaths & 33 & 13  & 17 \\
       \hline
    \end{tabular}
    \caption{Counts of computational and human judgement models that submitted predcitions of cases and deaths by the COVID-19 Forecast hub deadline stratified by survey~\label{tab.counts}}
\end{table}



\clearpage

\end{document}